\documentclass{iopart}

\usepackage{ucs}
\usepackage[utf8x]{inputenc}
\usepackage[english]{babel}
\usepackage{iopams} 
\usepackage{amssymb}
\usepackage[dvips]{graphicx}
\usepackage{psfrag}
\usepackage{tensor}
\usepackage{xspace}
\usepackage{wrapfig}

\usepackage[ps2pdf, bookmarks=true,
            bookmarksopen=true,colorlinks=false]{hyperref}

\hypersetup{%
  pdftitle = {Investigations of solutions of Einstein's field equations close to lambda-Taub-NUT},
  pdfsubject = {},
  pdfauthor = {Florian Beyer},
  pdfkeywords = {}%
}

\newcommand{\R}{\mathbb R}

\renewcommand{\Re}{\mathrm{Re}\,}
\renewcommand{\Im}{\mathrm{Im}\,}
\newcommand{\oplo}{\ensuremath{(1+1)}\xspace}
\newcommand{\tplo}{\ensuremath{(2+1)}\xspace}

\newcommand{\Stwo}{\ensuremath{\mathbb S^2}\xspace}
\newcommand{\SoXSt}{\ensuremath{\mathbb S^1\times\mathbb S^2}\xspace}
\renewcommand{\S}{\ensuremath{\mathbb S^3}\xspace}
\newcommand{\tildeS}{\ensuremath{\tilde{\mathbb S}^3}\xspace}
\newcommand{\T}{\ensuremath{\mathbb T^3}\xspace}
\newcommand{\U}{\ensuremath{\mathrm{U(1)}}\xspace}
\newcommand{\SU}{\ensuremath{\mathrm{SU(2)}}\xspace}

\newcommand{\normconstr}{\mathrm{Norm}^{\textrm{\small(\textit{constr})}}}

\newcommand{\Connection}[3]{\Gamma\indices{_{#1}^{#2}_{#3}}}

\newcommand{\scri}{\ensuremath{\mathcal J}\xspace}
\newcommand{\scrim}{\ensuremath{\mathcal J^-}\xspace}
\newcommand{\scrip}{\ensuremath{\mathcal J^+}\xspace}

\newcommand{\LTN}{$\lambda$-Taub-NUT\xspace}
\newcommand{\eps}{\ensuremath{\varepsilon}\xspace}

\newcommand{\Eqref}[1]{\eref{#1}}
\newcommand{\Eqsref}[1]{\eref{#1}}
\newcommand{\Sectionref}[1]{section~\ref{#1}}
\newcommand{\Figref}[1]{figure~\ref{#1}}

\newcommand{\term}[1]{#1}

\graphicspath{{figures/}}

\begin{document}

\bibliographystyle{hplain}

\title{Investigations of solutions of Einstein's field equations close
  to $\lambda$-Taub-NUT}

\author{Florian Beyer}

\address{KTH Matematik\\10044 Stockholm\\Sweden}

\ead{fbeyer@math.kth.se}

\begin{abstract}
  We present investigations of a class of solutions of Einstein's field
equations close to the family of \LTN spacetimes. The studies are done
using a numerical code introduced by the author elsewhere. One of the
main technical complication is due to the \S-topology of the Cauchy
surfaces.  Complementing these numerical results with heuristic
arguments, we are able to yield some first insights into the strong
cosmic censorship issue and the conjectures by Belinskii, Khalatnikov
and Lifschitz in this class of spacetimes. In particular, the current
investigations suggest that strong cosmic censorship holds in this
class. We further identify open issues in our current approach and
point to future research projects.

\end{abstract}


\maketitle

\section{Introduction}
Motivated by the desire to understand the history and fate of our
universe, studies of cosmological solutions of Einstein's field
equations have a long tradition.  Observations indicate that there was
a big bang in the distant past, and indeed, the simplest cosmological
models, namely the Friedmann solutions for reasonable matter fields,
predict such a collapse of spacetime. Now, the question arises whether
such curvature singularities occur generically in solutions of
Einstein's field equations or only under the strong symmetry
assumptions underlying the Friedmann models. This problem was solved,
namely due to the Hawking-Penrose singularity theorems \cite{hawking}
we can expect incompleteness of causal geodesics in a wide class of
situations. However, the information about the origin of the singular
behavior provided by these theorems is very limited. One would expect
that curvature singularities develop, as in the Friedmann case, which
'stop' the geodesics. But one was able to find solutions, as the \LTN
spacetimes discussed below, where the curvature stays bounded but the
spacetime becomes acausal in a well-defined sense.  Let us restrict in
all our investigations to vacuum with positive cosmological constant
$\lambda$, in which case Einstein's field equations (EFE) in geometric
units $c=1$ and $G=1/(8\pi)$ read
\begin{equation}
  \label{eq:EFE}
  R_{\mu\nu}=\lambda\, g_{\mu\nu},
\end{equation}
and to four spacetime dimensions.  Here, $g_{\mu\nu}$ is a Lorentzian
metric of signature $(-,+,+,+)$; $R_{\mu\nu}$ represents the Ricci
tensor of $g_{\mu\nu}$. One particularly important prototype solution
of these equations is the de-Sitter spacetime \cite{hawking}. The
positive cosmological constant is motivated by current observations
\cite{Sanchez06,Spergel06}. Under these (but actually under much
wider) assumptions, the Cauchy problem of EFE is well-posed and leads
to the notion of the maximal globally hyperbolic development (MGHD) of
a given Cauchy data set. See \cite{andersson04a} and references
therein for a review of the details. Now, for Cauchy data
corresponding to a \LTN spacetime for instance, the corresponding MGHD
is extendible in several inequivalent ways as smooth solutions of
EFE. In this sense, Einstein's theory has only limited predictive
power. The hope, expressed by strong cosmic censorship (SCC)
essentially first formulated by Penrose \cite{Penrose69,penrose1979},
is that, although examples like the \LTN spacetimes exist, the MGHD
corresponding to a \textit{generic} Cauchy data set is
inextendible. This would rescue the predictive power of Einstein's
theory. So far, there is neither a clear idea which topology on the
space of initial data is supposed to give rise to the notion of
'genericy' here, nor which class of extensions to choose in
general. Although this issue has only been solved successfully so far
in special situations, some of which are commented on below,
interesting new insights have been obtained as summarized in
\cite{andersson04a,Rendall05}.

Besides strong cosmic censorship there are other open problems of
interest for cosmological spacetimes.  Belinskii, Khalatnikov, and
Lifschitz \cite{lifshitz63,belinskii70,belinskii82} conjectured, what
we nowadays call BKL-conjecture, a description of the properties of
solutions with curvature singularities. We give only few further
details later, and just remark that it is still a matter of active
research whether their conjecture holds generically
\cite{andersson04a,Rendall05}.  Another outstanding problem is to find
a characterization of the development of generic cosmological
solutions in the expanding time direction, and the notion of cosmic
no-hair essentially introduced in \cite{Hawking82} is one promising
scenario.

In our work, we focus on the class of Gowdy spacetimes with spatial
\S-topology which yields one of the main technical complications. In
that class, the issue of SCC and the BKL-conjecture is still unsolved;
however, there are strong expectations due to results in the Gowdy
case with spatial \T-topology. We restrict to a special class of
solutions which obey a generalized notion of cosmic no-hair by
construction. Further, this class is close to \LTN in the sense that
it incorporates a ``perturbation'' parameter \eps such that $\eps=0$
corresponds to a \LTN spacetime. Our solutions are constructed using
Friedrich's notion of the \scrip-Cauchy problem discussed below by
means of a numerical code introduced in \cite{beyer08:code}.  The
purpose of our investigations is to check whether SCC is true in this
class and how the dynamics gives rise to it when \eps approaches
zero. We are further interested in the BKL-conjecture. We see these
investigations in this special class as first steps in order to shed
further light on the general SCC and BKL-conjectures.

The paper is organized as follows. First we give the relevant
background material in \Sectionref{sec:background}, including some
remarks about the conformal field equations and future asymptotically
de-Sitter spacetimes, the definition of the \LTN family, a short
introduction into the class of Gowdy spacetimes with spatial
\S-topology, our family of initial data and a few comments about our
numerical method. In \Sectionref{sec:results}, we write our main
result, support it by means of numerical computations and heuristic
arguments and discuss the current limitations of the approach. Finally
we conclude.


\section{Background material}
\label{sec:background}
\subsection{Notations and conventions}
In this paper, we will always assume Einstein's summation convention
when we write tensorial expressions. Any tensor can be represented by
the abstract symbol, say, $T$ or by abstract index notation, e.g.\
$T\indices{^\mu_\nu}$, dependent on the context. Note however, that
when we write such an indexed object, we can mean either the abstract
tensor or a special component with respect to some basis. Hopefully
this will be always clear from the context. Our convention is that
Greek indices $\mu,\nu,\ldots=0,\ldots,3$ refer to some choice of
local spacetime coordinates, whereas Latin indices
$i,j,\ldots=0,\ldots 3$ represent indices with respect to some
orthonormal frame field. When we have chosen a time coordinate $x^0$,
then Greek indices $\alpha,\beta,\ldots=1,2,3$ represent spatial
coordinates, and for a choice of frame $\{e_i\}$ with timelike vector
field $e_0$, Latin indices $a,b,\ldots=1,2,3$ refer to spatial frame
indices. Writing $\{e_i\}$ just means the collection of tangent vector
fields $\{e_0,\ldots,e_3\}$. If a $2$-indexed object is written in
brackets, for instance $(T\indices{^\mu_\nu})$, we mean the matrix of
its components, where the first index labels the lines of the matrix
and the second one the columns.

\subsection{Future asymptotically de-Sitter spacetimes and the
  conformal field equations}
Let a smooth Lorentzian manifold $(\tilde M,\tilde g_{\mu\nu})$ be
given; not necessarily a solution of EFE. According to Penrose
\cite{penrose1963,penrose1979}, it is said to have a smooth
\term{conformal compactification} if there exists a smooth Lorentz
manifold-with-boundary $(M,g)$ with boundary $\scri:=\partial M$ and a
smooth function $\Omega:M\rightarrow\R$ such that
\begin{enumerate}
\item there is a diffeomorphism 
  $\Phi:\tilde M\rightarrow M\backslash\scri$ such that
  $\tilde g=\Phi_*\left(\Omega^{-2} g\bigr|_{M\backslash\scri}\right)$,
\item we have $\Omega>0$ in the interior of $M$, and, $\Omega=0$ and
  $d\Omega\not=0$ on $\scri$.
\end{enumerate}
For simplicity we will always identify $\tilde M$ with $\Phi(\tilde
M)\subset M$. Under these assumptions, \scri is called conformal
boundary and can be interpreted as representing ``infinity'' of the
physical spacetime $(\tilde M,\tilde g_{\mu\nu})$.  In the following,
quantities which are related to the physical metric $\tilde
g_{\mu\nu}$ are marked with a tilde, while those related to the
conformal metric $g_{\mu\nu}$ are not marked. For a recent discussion
and several examples see \cite{Friedrich2002}. The main prototype
solution of EFE with $\lambda>0$ with a smooth conformal
compactification is the de-Sitter solution. Further examples are
provided by the \LTN family discussed below.  Now, in general if
$(\tilde M,\tilde g_{\mu\nu})$ is a solution of \Eqref{eq:EFE} in
vacuum with $\lambda>0$, then the corresponding conformal boundary
$\scri$ must be spacelike. In this case, we define \scrip as the
intersection of \scri with the chronological future of $\tilde M$ in
$M$ and analogously \scrim. We assume in all what follows that \scri
is the disjoint union of \scrip and \scrim, each of which is allowed
to be empty so far.  If \scrip is non-empty, we call $(\tilde M,\tilde
g_{\mu\nu})$ future asymptotically de-Sitter (FAdS); in the same way
one can define past asymptotically de-Sitter solutions. According to
\cite{galloway2002}, a FAdS spacetime with compact \scrip is globally
hyperbolic if and only if it is future asymptotically simple, i.e.\
all future directed inextendible null geodesics starting in $\tilde M$
have future endpoints on \scrip. Then, all Cauchy surfaces of
$(M,\tilde g_{\mu\nu})$ are actually homeomorphic to \scrip. From the
physical point of view it is further important that all FAdS solutions
obey a generalized notion of the cosmic no-hair picture
\cite{beyer:PhD}, and hence provide a model of the current
observational predictions for the future of our universe.

In order to compute solutions of the EFE in terms of conformal
variables directly, Friedrich succeeded in finding a formulation of
EFE called conformal field equations (CFE), which is both conformally
invariant and regular even on \scri. By conformal invariance we mean
that in addition to the coordinate gauge freedom of Einstein's theory,
the CFE obey the conformal gauge freedom; i.e.\ the possibility to
conformally rescale the conformal metric with a smooth positive
function leaving the physical metric invariant. By 'regular' we mean
that these equations provide evolution equations which are hyperbolic
even in \scri. A review of his results and further references are
given in \cite{Friedrich2002}. His formalism is restricted to four
spacetime dimensions, the relevant case for our
investigations. Anderson \cite{Anderson04} was able to find a
different formulation applicable in arbitrary even spacetime
dimensions. We, however, restrict here to Friedrich's CFE which we
write in a special form below. Under the assumptions above, in
particular for $\lambda>0$, the CFE allow one to formulate what we
call $\scrip$-Cauchy problem \cite{DeSitter}; Anderson's
generalization of this concept was proposed in \cite{Anderson04}. The
idea is to prescribe data for the CFE on the hypersurface $\scrip$
including its manifold structure subject to certain constraints
implied by the CFE. These data can be integrated into the past by
means of evolution equations implied by the CFE. The \scrip-Cauchy
problem is well-posed and the unique FAdS solution corresponding to a
given choice of smooth data on \scrip is smooth as long as it can be
extended into the past.  It is remarkable that this Cauchy problem
allows us to control the future asymptotics of the solutions
explicitly by deciding on the data on \scrip. Concerning the past
behavior, the concepts of MGHDs and SCC with respect to data on \scrip
are well defined, however, there is only limited a priori control in
the same way as it is the case for the Cauchy problem of EFE.  In this
paper, we will give no details on how the constraints implied on
\scrip look and say only briefly what the relevant data components
are, since we do not want to introduce all the geometric concepts
now. Instead we refer to \cite{DeSitter,beyer:PhD} where the details
have been carried out. In order to choose a complete set of smooth
data on \scrip, besides certain quantities related to the gauge, it
turns out to be sufficient to choose the differentiable manifold
structure of $\scrip$, which we assume to be the standard one on \S,
the induced conformal $3$-metric $h_{\alpha\beta}$ and the electric
part of the rescaled conformal Weyl tensor $E_{\alpha\beta}$
\cite{Friedrich2002,FriedrichNagy}. From those components the complete
data set is fixed in terms of algebraic or differential relations. The
choice of these components is unconstrained except for
$E_{\alpha\beta}$ which has to satisfy $D^\alpha
E_{\alpha\beta}=0$. Here $D$ is the Levi-Civita covariant derivative
operator of $h_{\alpha\beta}$, and the index in the constraint has
been raised by means of $h_{\alpha\beta}$. Below, we mention a special
class of solutions of this constraint.

For our purposes, we decided to use the so-called \term{general
  conformal field equations} which are the CFE in conformal Gauß gauge
\cite{AntiDeSitter,Friedrich2002}. This fixes the gauge freedom for
the coordinates, the conformal factor and a frame which is orthonormal
with respect to the conformal metric.  In our applications, we
specialize the gauge even further to what we call \term{Levi-Civita
  conformal Gauß gauge} \cite{beyer:PhD}. One assumption for that is
that the conformal geodesics representing the timelines leave $\scrip$
orthogonally. In particular this implies that they correspond to
physical geodesics up to parametrization \cite{ConfGeodesics}. Hence
this gauge can be considered as a (standard) Gauß gauge with respect
to the conformal (and also the physical) metric up to
parametrization. We can expect that this simple choice of gauge can
lead to serious problems in the evolution, and in fact, we will
discuss problems related to the gauge later. In any case, assuming
without loss of generality $\lambda=3$ and having fixed the residual
gauge initial data as described in \cite{beyer:PhD}, a hyperbolic
reduction of the general conformal field equations in Levi-Civita
conformal Gauß gauge is \numparts
\begin{eqnarray}
  &\partial_t e_a^c&=-\chi\indices{_a^b}e\indices{_b^c},\\
  &\partial_t\chi_{ab}&=-\chi\indices{_{a}^c}\chi_{cb}
  -\Omega E_{ab}
  +L\indices{_{{a}}_{b}},\\
  &\partial_t\Connection abc
  &=-\chi\indices{_a^d}\Connection dbc
  +\Omega B_{ad}\epsilon\indices{^b_c^d},\\
  &\partial_t L_{ab}
  &=-\partial_t\Omega\, E_{ab}-\chi\indices{_a^c}L_{cb},\\
  \label{eq:Bianchi1}
  &\partial_t E_{fe}-D_{e_c}B_{a(f}\epsilon\indices{^a^c_{e)}}
  &=-2\chi\indices{_c^c}E_{fe}
  +3\chi\indices{_{(e}^c}E_{f)c}
  -\chi\indices{_c^b}E\indices{_b^c}g_{ef},\\
  \label{eq:Bianchi2}
  &\partial_t B_{fe}+D_{e_c}E_{a(f}\epsilon\indices{^a^c_{e)}}
  &=-2\chi\indices{_c^c}B_{fe}
  +3\chi\indices{_{(e}^c}B_{f)c}
  -\chi\indices{_c^b}B\indices{_b^c}g_{ef},\\
  \label{eq:conffactor}
  &\Omega(t)&=\frac 12\, t\, (2-t),
\end{eqnarray}
\endnumparts
for the unknowns
\begin{equation*}
  u=\left(e\indices{_a^b}, \chi_{ab}, \Connection abc, L_{ab}, E_{fe},
    B_{fe}\right).
\end{equation*}
Among the unknowns are the spatial components $e\indices{_a^b}$ --
with respect to some reference basis specified below -- of a smooth
frame field $\{e_i\}$ which is orthonormal with respect to the
conformal metric and where, in this special gauge, $e_0=\partial_t$ is
the past directed timelike unit normal of the $t=const$-hypersurfaces,
the spatial frame components of the second fundamental form
$\chi_{ab}$ of the $t=const$-hypersurfaces with respect to $e_0$, the
spatial connection coefficients $\Connection abc$, given by
$\Connection abc e_b=\nabla_{e_a}e_c-\chi_{ac}e_0$ where $\nabla$ is
the Levi-Civita covariant derivative operator of the conformal metric,
the spatial frame components of the Schouton tensor $L_{ab}$, which is
related to the Ricci tensor of the conformal metric by
\begin{equation*}
L_{\mu\nu}=R_{\mu\nu}/2
  -g_{\mu\nu}g^{\rho\sigma}R_{\rho\sigma}/12,
\end{equation*}
and the spatial frame components of the electric and magnetic parts of
the rescaled conformal Weyl tensor $E_{ab}$ and $B_{ab}$ defined with
respect to $e_0$.  In this special conformal Gauß gauge, the timelike
frame field $e_0$ is hypersurface orthogonal, i.e.\ $\chi_{ab}$ is
symmetric.  In order to avoid confusions, we point out that the
conformal factor $\Omega$ is part of the unknowns in Friedrich's
formulation of the CFE. However, in the special case of vacuum with
$\lambda>0$, it is possible to integrate its evolution equation in any
conformal Gauß gauge explicitly \cite{AntiDeSitter} so that it takes
the special form \Eqref{eq:conffactor} in our gauge.  Hence, if the
solution develops a smooth \scrim then it must correspond to the
$t=2$-hypersurface. The initial hypersurface $\scrip$ corresponds to
$t=0$, and $t$ increases into the past. We note further that, since
$E_{ab}$ and $B_{ab}$ are tracefree, we identify
$E_{33}=-E_{11}-E_{22}$; the same for the magnetic part.  The
evolution equations of $E_{ab}$ and $B_{ab}$ are derived from the
Bianchi system \cite{Friedrich2002}. In our gauge, the constraint
equations implied by the Bianchi system take the form
\begin{equation}
  \label{eq:bianchi_constraints}
  D_{e_c} E\indices{^c_e}
  -\eta\indices{^a^b_e}B_{da}\chi\indices{_b^d}=0,\quad
  D_{e_c} B\indices{^c_e}
  +\eta\indices{^a^b_e}E_{da}\chi\indices{_b^d}=0,
\end{equation}
where $\eta$ is the volume form of the spatial conformal metric with
$\eta_{123}=1$, and indices are shifted by means of the conformal
metric.  The other constraints of the system above are equally
important but are ignored for the presentation here. Further
discussions of that evolution system and the quantities involved can
be found in the references above.

As the reference basis tangential to the $t=const$-hypersurfaces
assumed to be \S topologically, we choose frame vector fields
$\{Y_a\}$ which are \SU-left invariant and satisfy the commutator
relations
\begin{equation*}
  {}[Y_a,Y_b]=2\sum_{c=1}^3\eta\indices{_a_b_c}Y_c,
\end{equation*}
where $\eta\indices{_a_b_c}$ is the totally antisymmetric symbol with
$\eta\indices{_1_2_3}=1$. These requirements fix those vector fields
uniquely up to the specific choice of realization of the \SU-action on
\S and up to certain transformations on the three-dimensional real
vector space of \SU-left invariant vector fields. We do not list our
choices involved here but rather write the explicit expressions
below\footnote{To be more precise we choose $\{Y_a\}$ as the reference
  frame on the spacetime as follows. Let $\{Y_a\}$ be given as before
  on one $t=const$ Cauchy surface. Then we extend the frame to the
  spacetime, keeping the same symbol $\{Y_a\}$, by means of the
  conditions $[\partial_t,Y_a]=0$.}. Henceforth we write
$e_a=e\indices{_a^b}Y_b$, and in the Bianchi system
\Eqsref{eq:Bianchi1} and \Eqref{eq:Bianchi2} and corresponding
constraints \Eqref{eq:bianchi_constraints}, the orthonormal frame
fields are interpreted as linear differential operators whose
coefficients are given by $e\indices{_a^b}$ and by the components of
$\{Y_a\}$ in some coordinate system which we fix in a moment.

We finally note how the Kretschmann scalar of the physical metric can
be computed in terms of the variables above if the physical metric is
a solution of EFE in vacuum with $\lambda>0$,
\begin{equation}
  \label{eq:physKretschConfQuant}
  \tilde \kappa=\tilde R^{ijkl}\tilde R_{ijkl}
  =24 \frac{\lambda^2}9 +8\Omega^6\sum_{a,b=1}^3 
  \left(|E_{ab}|^2-|B_{ab}|^2\right).
\end{equation}

\subsection{\texorpdfstring{\LTN}{lambda-Taub-NUT} solutions}
\label{sec:LTaubNUT}
Consider the function $V:\R\rightarrow\R$ defined by
\[V(\tau)=D_0\tau^4 +2(3D_0-2)\tau^2+C_0\tau+4-3D_0\] where
$(D_0,C_0)\in\R^{>0}\times \R$ are arbitrary parameters. For given
$(D_0,C_0)$, let $I$ be a maximal interval on which
$V>0$. Corresponding to the choice of cosmological constant
$\lambda>0$, parameters $(D_0,C_0)$ and interval $I$, we define the
$\lambda$-Taub spacetime \cite{Brill1978} as the Lorentz manifold
$(\tilde M, \tilde g)$ with $\tilde M=I\times\S$ and
\begin{eqnarray*}
  \tilde g&=\frac 3\lambda D_0\Bigl[&
  -\frac{1+\tau^2}{V(\tau)}d\tau^2
  +\frac{V(\tau)}{1+\tau^2}\,\sigma^3\otimes\sigma^3\\
  &&+(1+\tau^2)(\sigma^1\otimes\sigma^1+\sigma^2\otimes\sigma^2)
  \Bigr]
\end{eqnarray*}
which can be shown to be a solution of $\tilde
R_{\mu\nu}=\lambda\tilde g_{\mu\nu}$.  Here,
$\{\sigma^a\}$ is the dual of the frame $\{Y_a\}$. Note that for a
given choice of $(D_0,C_0)$ there can exist several distinct choices
of the interval $I$ and hence several distinct $\lambda$-Taub
spacetimes.  

\psfrag{time}[tc][tc][0.7]{time coordinate $\tau$}
\psfrag{J+}[bl][tc][0.7]{\scrip}
\begin{wrapfigure}{l}{0.4\linewidth}
  \centering
   \includegraphics[width=0.39\textwidth]{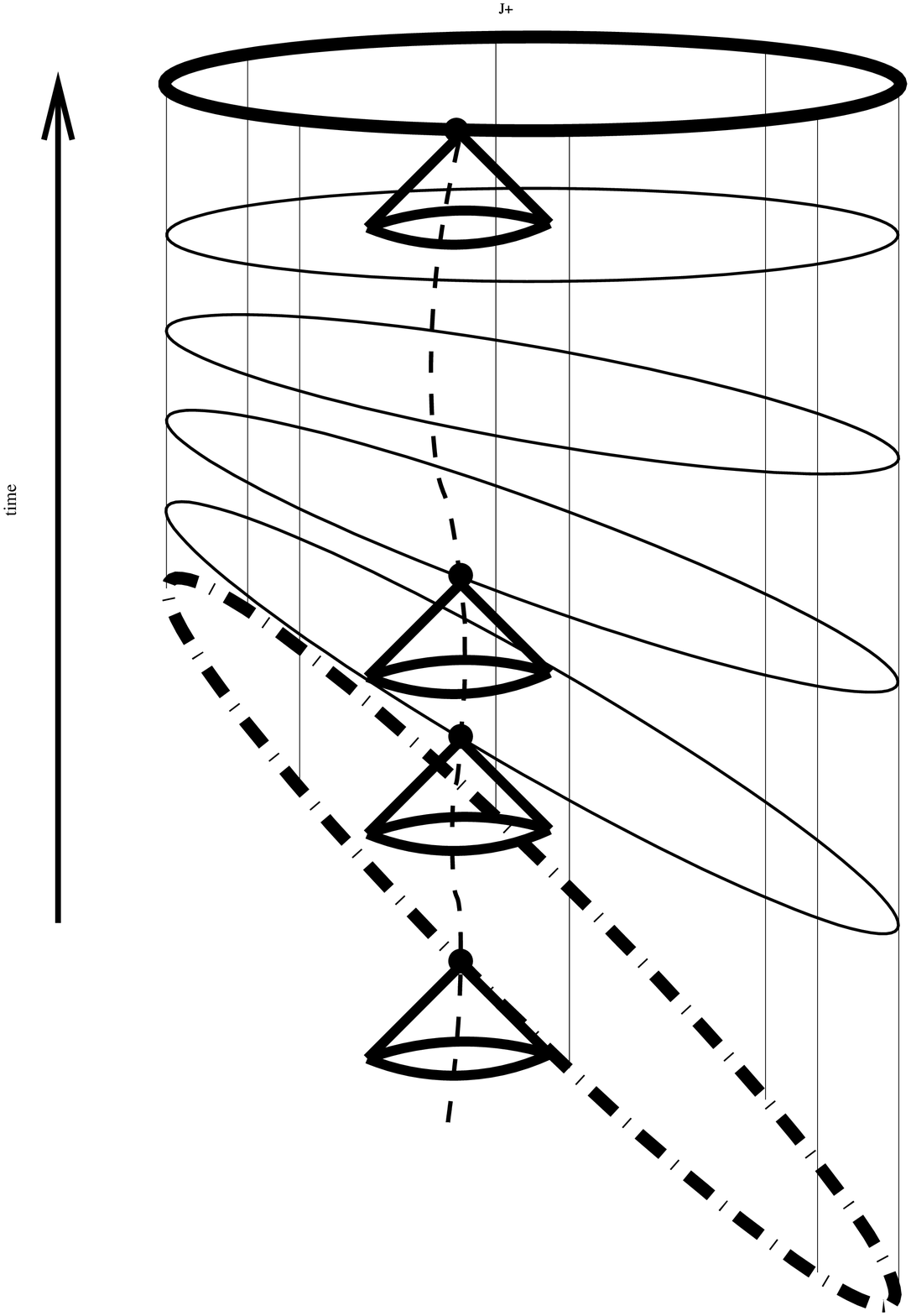}
  \caption{Causal structure of a singular $\lambda$-Taub solution.}
  \label{fig:extendibleTaub}
\end{wrapfigure}
The $\lambda$-Taub spacetimes are globally hyperbolic and maximal with
this property. The de-Sitter solution is recovered by choosing the
parameters $(D_0,C_0)=(\lambda/3,0)$, in which case $I=\R$. Indeed,
due to the nonlinear stability of the de-Sitter spacetime
\cite{friedrich86,Anderson04}, there is an open neighborhood $U$ of
the de-Sitter point in the parameter space $\R^{>0}\times\R$ in which
all corresponding $\lambda$-Taub spacetimes are causal geodesically
complete and asymptotically de-Sitter in the future and in the
past. In general, for any given choice of $(D_0,C_0)$, one can find
$\tau_0\in \R\cup\{-\infty\}$ such that one pick $I=]\tau_0,\infty[$,
and we will assume such $I$ in the following. The corresponding
$\lambda$-Taub solution is FAdS, as depicted in
\Figref{fig:extendibleTaub}; note that in our convention
$\partial_\tau$ is future directed.  Now, if $V$ has zeros on $\R$,
i.e.\ if $\tau_0\not=-\infty$, we will say that the corresponding
$\lambda$-Taub spacetime is singular. Indeed, the metric is clearly
singular when $V$ becomes zero, however, curvature invariants are
bounded. One can show analogously to the case treated in
\cite{chrusciel93} that singular $\lambda$-Taub spacetimes are always,
independent of the type of zero of $V$ at $\tau_0$, extendible and
causal geodesically incomplete. In particular, this provides counter
examples \cite{Anderson08} to theorem~1.3 in \cite{Anderson04}.  Any
extension of a singular $\lambda$-Taub solution is non-globally
hyperbolic and will be referred to as \LTN.  We will not write
extensions here but only note the following. The function $V$ can only
have a single or double real zero at $\tau=\tau_0$. In both cases the
$\tau=\tau_0$-hypersurface corresponds to a smooth Cauchy horizon
diffeomorphic to \S ruled by closed null curves.  In the first case we
will call the Cauchy horizon ``non-degenerate'', in the second case
``degenerate''. This corresponds to the terminology of
\cite{moncrief83} which can be seen by constructing extensions in
``Gaussian null coordinates''.  \Figref{fig:extendibleTaub} depicts a
conformal diagram of a singular $\lambda$-Taub solution. In that
figure, circles correspond to $\tau=const$-hypersurfaces, the bold
circle represents \scrip and the dashed circle the Cauchy horizon.  We
will be particularly interested in the non-degenerate case here.  Note
that we will often abuse terminology and speak, instead of
$\lambda$-Taub, often of \LTN.

Singular $\lambda$-Taub solutions seem to be counter examples to
SCC. Within the Bianchi IX class for $\lambda=0$, the analogous
Taub-NUT spaces \cite{Taub51,NUT63} for $\lambda=0$ have been shown to
be non-generic in a well-defined sense \cite{Ringstrom99}. For
$\lambda>0$ one believes that the same holds, but no such rigorous
result has been found yet. A more general extensively studied symmetry
class is Gowdy symmetry which we discuss in the following
subsection. In general, further relevant rigorous results concerning
SCC for solutions with spatial \S-topology are
\cite{moncrief83,Friedrich98,moncrief84}.

\subsection{Coordinates and Gowdy spacetimes}
\label{sec:gowdy}
Gowdy symmetry \cite{Gowdy73} is defined by the presence of a smooth
effective isometric action of the group $\U\times\U$ with spacelike
orbits. As summarized in \cite{chrusciel1990}, if we restrict to
smooth orientable connected compact spatial 3-manifolds, the only
admitted spatial topologies are \T, \SoXSt and \S (and lens spaces
which are always included in the following discussion without further
comment). Any two of such group actions on one of these topologies are
equivalent in the sense of \cite{chrusciel1990}.

For the latest state of knowledge in the class of \T-Gowdy spacetimes,
we refer to \cite{Ringstrom06,Ringstrom06b} and references therein; we
rather focus on the \S-case here. A representation of the action of
the Gowdy group on \S is given by
\begin{equation}
  \label{eq:eulerangleparm}
  \eqalign{x_1&=\cos\frac\chi 2\cos\lambda_1, 
  \quad x_2=\cos\frac\chi 2\sin\lambda_1,\cr
  x_3&=\sin\frac\chi 2\cos\lambda_2, 
  \quad x_4=\sin\frac\chi 2\sin\lambda_2,}
\end{equation}
where we assume the standard embedding of \S into $\R^4$.  The
parameter $\chi\in[0,\pi]$ labels the group orbits while $\lambda_1$
and $\lambda_2$, both defined on $\R\,\mathrm{mod}\,2\pi$, can be
considered as the group parameters. The action degenerates exactly
when $\chi=0$ or $\chi=\pi$, i.e.\ the group orbits become one
dimensional. These submanifolds of $\mathbb{S}^1$-topology are
referred to as symmetry axes in the following. By abusing the
terminology slightly, we often also speak of the south pole when we
mean the symmetry axis given by $\chi=\pi$; analogously, by north pole
we mean the axis given by $\chi=0$.
Equation~\Eqref{eq:eulerangleparm} can also be considered as the
defining equation for a coordinate patch on \S; these coordinates
$(\chi,\lambda_1,\lambda_2)$ are called Euler angle
parametrization. We will however rather use the coordinates
$(\chi,\rho_1,\rho_2)$ determined by
\begin{equation*}
  \lambda_1=(\rho_1+\rho_2)/2,\quad\lambda_2=(\rho_1-\rho_2)/2.
\end{equation*}
The Euler angle parametrization covers smoothly the dense subset of
$\tildeS\subset \S$ which is \S minus those points given by 
$\chi=0$ or $\chi=\pi$ where the coordinates become
singular. Our choice of representation of the frame $\{Y_a\}$ yields
the following expressions
\numparts
\begin{eqnarray}
  \label{eq:coordinate_repr_standard_frame_Y1}
  Y_1=2\sin \rho_1\,\partial_\chi
  +2\cos \rho_1\,
  \left(\cot\chi\,\partial_{\rho_1}-\csc\chi\,\partial_{\rho_2}\right),\\
  Y_2=2\cos \rho_1\,\partial_\chi
  -2\sin \rho_1\,
  \left(\cot\chi\,\partial_{\rho_1}-\csc\chi\,\partial_{\rho_2}\right),\\
  \label{eq:coordinate_repr_standard_frame_Y3}
  Y_3=2\partial_{\rho_1}.
\end{eqnarray}
\endnumparts

Now, on a Gowdy-invariant spacetime with spatial \S-topology let us
choose spacetime coordinates $(t,\chi,\lambda_1,\lambda_2)$ such that
the $t=const$-hypersurfaces are Cauchy surfaces, and such that on a
$t=t_0$-hypersurface $\Sigma_0$ the Gowdy orbits are
$\chi=const$-surfaces parametrized canonically as above by $\lambda_1$
and $\lambda_2$. On $\Sigma_0$, a Killing algebra basis is
$\{\partial_{\lambda_1},\partial_{\lambda_2}\}$, but also
$\{Y_3,Z_3\}$ where $\{Z_a\}$ are \SU-right invariant vector fields
defined in the same way as $\{Y_a\}$ before; in particular
$Z_3=2\partial_{\rho_2}$.  Let us now extend the fields $Y_3$ and
$Z_3$ from $\Sigma_0$ to the spacetime by means of the conditions
$[\partial_t,Y_3]=[\partial_t,Z_3]=0$. Then it is clear directly that
$\{Y_3,Z_3\}$ is a basis of the Killing algebra at every instance of
time if and only if this is the case for
$\{\partial_{\lambda_1},\partial_{\lambda_2}\}$. It is obvious that
the latter does not need to be fulfilled for all foliations; however,
it is true in our gauge \cite{beyer:PhD} and also in areal gauge
defined below.

The issues of SCC and the BKL-conjecture in the \S-Gowdy class have so
far only been solved in the polarized vacuum case with $\lambda=0$
\cite{Isenberg89,Chrusciel90}. In particular, there is a nice
characterization of those non-generic Gowdy-invariant data whose MGHD
is extendible through a Cauchy horizon. For the non-polarized case,
both issues are open. From the results obtained in the \T-case, one
expects that spikes are present generically at the singularity, but in
particular the behavior at the symmetry axes is not completely
understood \cite{Stahl02}. The purpose of our investigations is to
shed further light on SCC and the issue of spikes for \S-Gowdy
perturbations of the \LTN family. Similar numerical investigations in
the \SoXSt-case have been carried out in \cite{garfinkle1999}.

Let us introduce the notion of the orbit area density. This quantity
is defined as $\sqrt{\det
  (g(\partial_{\lambda_a},\partial_{\lambda_b}))}$ ($a=1,2$). In the
\S-case one can show that it is proportional to $\sin\chi$ and we will
speak of the rescaled orbit area density when this factor has been
divided out. For $\lambda=0$ in vacuum, Chru\'{s}ciel
\cite{chrusciel1990} succeeded in showing that an open subset of the
MGHD corresponding to generic Gowdy-invariant Cauchy data can be
covered by an areal foliation, i.e.\ a foliation of spacelike
hypersurfaces given by the level sets of the rescaled orbit area
density, and that the orbits reach zero area either at the boundary of
the MGHD or in the interior. In the polarized case, the result is
stronger, namely, that the boundary of the MGHD is attained exactly
where the orbit area vanishes in the generic case. There are no such
results for $\lambda>0$. In particular, it cannot be expected that
solutions which collapse and later re-expand due to the repelling
force of the cosmological constant can be covered completely by areal
foliations.

Other aspects of Gowdy symmetry and the more general \U-symmetry which
are needed to develop our numerical method are summarized in
\cite{beyer08:code}. In particular, it is explained there how our
formulation of the equations can be reduced to \oplo dimensions in the
Gowdy case in an indirect manner, since it is not possible
directly. We should point out that this reduction allows us to
restrict to the coordinate locations $\rho_1=0$ and an arbitrary
(fixed) value of $\rho_2$. Hence effectively the unknowns depend only
on $(t,\chi)$ and our problem reduces to $1+1$ in the Gowdy case. All
our computations here take place at $\rho_1=0$; the solution at other
points can be computed from the formulae in \cite{beyer08:code}
straight forwardly.

\subsection{Class of initial data}
\label{sec:id}
As initial data on \scrip for the CFE in Levi-Civita conformal Gauß
gauge, we use the data derived for \U- and Gowdy symmetry in
\cite{beyer:PhD}, see also \cite{beyer08:code}. Here we restrict to
Gowdy symmetry. We make no claims about the genericy of these data
even in the class of Gowdy data. Under the conventions above, the
first part of the data take following form \numparts
\begin{eqnarray}
  \label{eq:IDframe}
  (e\indices{_a^b})&=&\mathrm{diag}(1,1,a),\\
  (\chi_{ab})&=&\mathrm{diag}(-1,-1,-1),\\
  \Gamma\indices{_1^1_2}&=&0,\quad
  \Gamma\indices{_1^2_3}=-1/a,\quad
  \Gamma\indices{_2^1_2}=0, \quad
  \Gamma\indices{_2^1_3}=1/a,\\
  \Gamma\indices{_2^2_3}&=&0,\quad
  \Gamma\indices{_3^1_2}=1/a-2a,\quad
  \Gamma\indices{_3^1_3}=0,\quad
  \Gamma\indices{_3^2_4}=0,\\
  (L_{ab})&=&\mathrm{diag}\Bigl((5-3/a^2)/2,\,\,(5-3/a^2)/2,\,\,
  (-3+5/a^2)/2\Bigr),\\
  (B_{ab})&=&\mathrm{diag}\Bigl(-4(1-a^2)/a^3,
  \,\,-4(1-a^2)/a^3,\,\,8(1-a^2)/a^3\Bigr).
\end{eqnarray}
The induced conformal $3$-metric of \scrip is a Berger sphere with
anisotropy parameter $a>0$ and an adapted orthonormal frame which is
expressed by \Eqref{eq:IDframe}. Now, the only inhomogeneous part of
the initial data is given by the components of $E_{ab}$. In order to
construct solutions of the constraint $D^a E_{ab}=0$ on the background
defined by the expressions above, we expand the components of $E_{ab}$
in terms of spin-weighted spherical harmonics on \S, which under the
assumption of \U-symmetry are related to spherical harmonics $w_{np}$
on \Stwo. For our definition of the functions $w_{np}$ consult
\cite{beyer:PhD}. We restrict to solutions with $n=0$ and $n=2$. Then
the expressions are
\begin{equation*}
  (E_{ab})=\left(
    \begin{array}{ccc}
      E_0+\eps\,w_{20} & 0 & -\sqrt{2}\,a\, \eps\,\Re w_{21}\\
      0 & E_0+\eps\,w_{20} & -\sqrt{2}\,a\, \eps\,\Im w_{21}\\
      -\sqrt{2}\,a\, \eps\,\Re w_{21} 
      & -\sqrt{2}\,a\, \eps\,\Im w_{21} & -2(E_0+\eps\,w_{20})
    \end{array}\right).
\end{equation*}
\endnumparts
With respect to the Euler angle coordinates at $\rho_1=0$ we have
\[w_{20}=\cos \chi,\quad w_{21}=\sin \chi/\sqrt{2}.\] The parameter
$E_0$ is associated with the part of the data given by $n=0$, while
\eps parametrizes the inhomogeneous $n=2$ part.

Note that the solution corresponding to a general choice of parameters
in this family is of unpolarized Gowdy type. This has already been
claimed in \cite{beyer:PhD}; however, there the argument turned out to
be incomplete. In general, we call two initial data sets equivalent if
the corresponding MGHDs are isometric. In particular, two initial data
sets on \scrip are equivalent if there is a diffeomorphism from \scrip
to itself such that the pullback of all data fields equals the
corresponding fields in the other data set, up to gauge
transformations. In our special class of initial data now we find that
two initial data sets, one given by $\eps_1$ and the other by
$\eps_2=-\eps_1$, both with the same $a>0$ and $E_0\in\R$, are
equivalent by means of the diffeomorphism from \S to itself
represented by $(x_1,x_2,x_3,x_4)\mapsto (x_3,x_4,x_1,x_2)$. In terms
of the Euler coordinates, this diffeomorphism takes the form
$(\chi,\rho_1,\rho_2)\mapsto (\pi-\chi,\rho_1,-\rho_2)$.  Hence
without loss of generality we can assume that $\eps\ge 0$.  For
$\eps=0$, the corresponding MGHD of $\scrip$ is a $\lambda$-Taub
spacetime whose parameters $(D_0,C_0)$ can be identified with our
initial data parameters by
\[a=1/\sqrt{D_0},\quad E_0=C_0\sqrt{D_0}/2,\] for $\lambda=3$. In
Levi-Civita conformal Gauß gauge, all $\tau$-level sets are also
$t$-level sets. Note however that our convention is that
$\partial_\tau$ is future directed while $\partial_t$ is past directed
so that \scrip is given by $t=0$ and $\tau\rightarrow +\infty$,
respectively.

We note that in our formulation of the equations, choice of gauge and
with the choice of data above, the following ``boundary conditions''
for the fields $E_{ab}$ and $B_{ab}$ are implied at each instant of
time \cite{beyer08:code}. Introducing the new fields
\[E_{11}^*:=(E_{11}+E_{22})/2, \quad E_{22}^*:=(E_{11}-E_{22})/2,\]
and similar for the magnetic part $B_{ab}$, we find on both symmetry
axes at each instant of time
\begin{equation}
  \label{eq:BCs}
  E_{12}=E_{13}=E_{22}^*=E_{23}=0, \quad B_{12}=B_{13}=B_{22}^*=B_{23}=0,
\end{equation}
whereas $E_{11}^*$ and $B_{11}^*$ are not restricted. Similar
properties are obeyed by all symmetric Gowdy-invariant covariant
2-tensor fields, for instance $\chi_{ab}$. Also, one can derive
analogous conditions for some of the connection coefficients which we
do not write here.  In addition, the constraints of the Bianchi system
\Eqref{eq:bianchi_constraints} imply on both axes
\begin{equation}
  \label{eq:BCs2}
  \partial_{\chi} E_{11}^*=\partial_{\chi} E_{22}^*
  =\partial_{\chi} B_{11}^*=\partial_{\chi} B_{22}^*=0.
\end{equation}

\subsection{Numerical approach}
The numerical method which we apply in this paper will not be
discussed here, see \cite{beyer08:code} for that. In summary, it is a
pseudospectral single patch approach; as coordinates on \S we decided
for the Euler angle parametrization. The relevant singular terms are
then determined by the coordinate components of the reference frame
$\{Y_a\}$ in \Eqref{eq:coordinate_repr_standard_frame_Y1}-\Eqref{eq:coordinate_repr_standard_frame_Y3}. We succeeded in
regularizing spectrally the formally singular terms in the equations
\cite{beyer08:code}.

For our investigations here we made the following special choices
which have not yet been fixed in \cite{beyer08:code}. First, we
decided to use no automatic adaption for the spatial
resolution. Namely, it turned out to be more convenient to control the
spatial resolution manually in order to study convergence of the code
at very late times. As the time integrator we decided for the adaptive
fifth order ``embedded'' Runge Kutta scheme \cite{numericalrecipes}.
Further, we found that it is necessary to enforce the boundary
conditions \Eqref{eq:BCs} at each time step in order to keep the code
stable at very late times of the evolution; more details about our
``partial enforcement'' scheme can be found in
\cite{beyer08:code}. Last but not least we mention that we did not use
the quad-precision option of the Intel compiler \cite{Intel}. Except
for very small values of \eps, see more comments later,
double-precision turned out to be sufficient and the constraint growth
was more or less under control.


\section{Results and their analysis}
\label{sec:results}
\psfrag{-log(0.69599284161-t)}[tc][tc][0.7]{$-\ln(0.69599284-t)$}
\psfrag{chi/pi}[tc][tc][0.7]{$\chi/\pi$}
\psfrag{t}[tc][tc][0.7]{$t$}
\psfrag{Kretschmann L1}[cm][tl][0.7]%
{$\quad\quad\left\|\mathrm{Kretschmann}-24\right\|_{L_1}/16$}
\psfrag{Components of rescaled Weyl tensor at t=0.69598373}%
[cl][tl][0.7]{Rescaled Weyl tensor components}
\psfrag{Components of rescaled Weyl tensor at t=0.69599118}%
[cl][tl][0.7]{Rescaled Weyl tensor components}
\psfrag{Components of rescaled Weyl tensor at t=0.69599176}%
[cl][tl][0.7]{Rescaled Weyl tensor components}
\psfrag{Components of rescaled Weyl tensor at t=0.695992057}%
[cl][tl][0.7]{Rescaled Weyl tensor components} 
\psfrag{Components of rescaled Weyl tensor at t=0.69520300}%
[cl][tl][0.7]{Rescaled Weyl tensor components}
\psfrag{Components of rescaled Weyl tensor at t=0.69520491525}%
[cl][tl][0.7]{Rescaled Weyl tensor components}
\psfrag{4 Sin2chi Omega4 OVol2 at t=0.69520492900}%
[cl][tl][0.7]{$4\sin^{-2}\chi\,\,\Omega^4(t)(\mathrm{Orbit\,volume})^2$}
\psfrag{4 Sinm2chi Omega4 OVol2}[cm][tl][0.7]%
{$\quad\quad\quad\quad\quad\quad\quad 4\sin^{-2}\chi\,\,\Omega^4(t)(\mathrm{Orbit\,volume})^2$}
\psfrag{E11* at chi=0}%
[cl][tl][0.7]{$E_{11}^*$ at $\chi=0$}
\psfrag{Bianchi constraint norm}%
[cl][tl][0.7]{$\normconstr$}

\setcounter{totalnumber}{1}

\subsection{Setup and main results}
\label{sec:mainresults}
Our strategy to analyze the properties of solutions close to the \LTN
family is as follows. For fixed \LTN background spacetimes, i.e.\ for
given values of $a$ and $E_0$, we compute and analyze the solutions
corresponding to initial data given by several small values of
$\eps\ge 0$. In \cite{beyer:PhD} we have speculated that \LTN Cauchy
horizons might be stable for small $\eps$ in the sense that a smooth
Cauchy horizon eventually forms for all $0\le\eps<\eps_{crit}$ for
some $\eps_{crit}>0$. The analysis there was done on the basis of a
numerical code, namely the \tplo-code in \cite{beyer08:code}, which
was not well adapted to the demands of the problem. Here we report on
results obtained with the new \oplo-code introduced in
\cite{beyer08:code} which allows much more reliable
investigations. The main reason is that with the \oplo-code no spatial
resolution is wasted in one of the symmetry directions; recall however
that a direct reduction to \oplo in our formulation is not
possible. We still use the same gauge and formulation of the equations
as in the \tplo-code.

Let us choose the parameters $a$ and $E_0$ such that the \LTN solution
corresponding to $\eps=0$ has a non-degenerate Cauchy horizon, and let
$t_{TN}$ be the corresponding critical coordinate time in Levi-Civita
conformal Gauß gauge.  Our investigations suggest the following
results which supersede the previous speculations.
\begin{enumerate}
\item For each $\eps>0$, there is a time $0<t_\eps<t_{TN}$ when the
  curvature blows up; further $t_\eps\nearrow t_{TN}$ for
  $\eps\searrow 0$.
\item The mechanisms driving the blow up are ``qualitatively
  independent'' of \eps (in the sense below); hence blow up takes
  place always ``in the same way'' in this family.
\item The standard continuity property holds: the smaller $\eps>0$,
  the longer the corresponding solutions stays close to the
  \LTN solution given by $\eps=0$.
\end{enumerate}
These results suggest that SCC is true in this class of
solutions. Below we give more details on the dynamical processes which
seem to give rise to SCC.

In all of our numerical investigations below we choose $a=0.7$ and
$E_0=0$, and the reason is the following. Our aim is to study
perturbations of spacetimes with a Cauchy horizon, and this is the
case for all \LTN solutions with, for instance, $E_0=0$, and
$a\le\sqrt{3}/2\approx 0.866$ (for $\lambda=3$).  We expect that
qualitatively the results do not depend strongly on these parameters
if the parameters stay in a regime in which the background Cauchy
horizon is non-degenerate. This is the case e.g.\ for all
$a<\sqrt{3}/2$ and $E_0=0$. The situation might of course be very
different when the background is of degenerate type, for instance for
$a=\sqrt{3}/2$ and $E_0=0$, however, we do not study that case
here. For $a=0.7$, we have $t_{TN}\approx 0.69599284$.

\setcounter{footnote}{4}
\begin{figure}[t]
  \centering
  \includegraphics[width=0.5\textwidth]{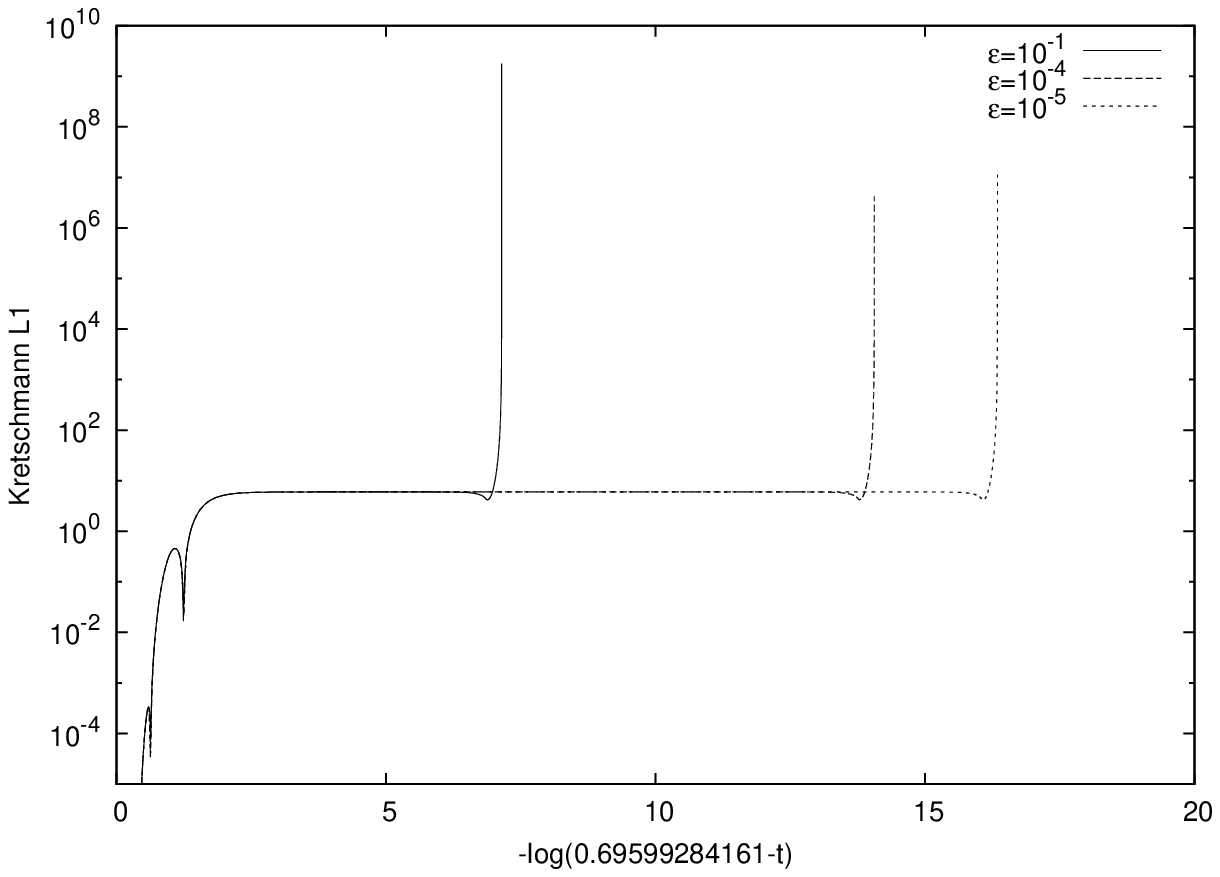}
  \caption{Curvature blow up for various values of \eps}
  \label{fig:kretschm_eps} 
\end{figure}
\Figref{fig:kretschm_eps} shows the numerical plot of a spatial
norm\footnote{This norm is the $L^1$-norm on \S, however, as mentioned
  in \cite{beyer08:code}, the volume element of the Euler angle
  coordinates is left out in the integral.}  of the physical
Kretschmann scalar versus time for various values of \eps. In that plot,
we have normalized the time axis such that the critical time $t_{TN}$
of the unperturbed solution is at roughly $20$. The smallest value of
\eps in our numerical series was $10^{-5}$, see comments on this
below. Note that more details on the numerical properties of the
solutions, at least in the case $\eps=10^{-1}$, are available in
\cite{beyer08:code}.

\subsection{Heuristic discussions of dynamical effects}
\subsubsection{Preparations}
Now we discuss the numerical results in a more detailed way and
complement them with heuristic arguments in order to give some
explanations and extrapolations of our results. A rigorous discussion
would be clearly preferable, however, is believed to be extremely
difficult since many terms with their individual sign structures seem
to be significant for the dynamics. One advantage of a numerical
approach is that it can be expected to provide information in
particular on intermediate evolution times. This might shed light on
for instance the question which dynamical processes are responsible to
implement SCC. The idea for the following heuristic discussion is to
consider the linearization of the evolution equation for small
$\eps>0$ around the \LTN solution given by $\eps=0$. One can believe
that this yields a good approximation for small \eps and small $t$.
Unexpectedly, we find below that a discussion exclusively based on the
linearized equations already fails at early times, i.e.\ in a small
past neighborhood of $\scrip$. However, by taking into account some
nonlinear effects supplementing the linearized equations, we are able
to obtain some insights. It should be noted that we are only
interested in qualitative arguments now and that none of the
simplifying assumptions or approximations considered in the following
are used for the numerical calculations. We point out again, that our
time coordinate which we always refer to runs backward in time, i.e.\
when we speak, e.g.\, of the early phase of the evolution we mean the
dynamics close to \scrip, hence very far in the future from the
physical point of view. After these discussions, we also elaborate on
the limitations of the current results and hence what remains to be
done in future work.

In order to proceed with the linearization we write for the unknowns
\[e\indices{_a^b}={\mathring{e}}\indices{_a^b} +\eps
{\hat{e}}\indices{_a^b},\quad
\chi\indices{_a_b}={\mathring{\chi}}\indices{_a_b} +\eps
{\hat{\chi}}\indices{_a_b},\quad
\Gamma\indices{_a^b_c}={\mathring{\Gamma}}\indices{_a^b_c} +\eps
{\hat{\Gamma}}\indices{_a^b_c},\quad \ldots,
\]
where higher order terms in \eps, which are in principle present for
some of these unknowns, have already been neglected. Below we express
some of the background \LTN components, marked by $\circ$, by means of
functions $a_1(t)$ and $a_3(t)$ and their time derivatives such that
\begin{eqnarray*}
  ({\mathring{e}}\indices{_a^b})(t)&=\mathrm{diag}(a_1(t),a_1(t),a_3(t)),\\
  (\mathring\chi_{ab})(t) &= 
  \mathrm{diag}(-a_1^\prime(t)/a_1(t),-a_1^\prime(t)/a_1(t),
  -a_3^\prime(t)/a_3(t)),\\
  {\mathring{\Gamma}}\indices{_3^1_2}(t)&=a_1(t)^2/a_3(t)-2a_3(t),\quad
  {\mathring{\Gamma}}\indices{_1^2_3}=
  -{\mathring{\Gamma}}\indices{_2^1_3}=-a_1(t)^2/a_3(t);
\end{eqnarray*}
all other connection coefficients of the background vanish. Here, a
prime denotes the derivative with respect to the time $t$ of the \LTN
solution given by $\eps=0$. The function $a_3(t)$ equals the initial
data parameter $a$ before at $t=0$, further $a_1(0)=1$. The explicit
form of the functions $a_1(t)$ and $a_3(t)$ can be derived from the
relations given in \Sectionref{sec:LTaubNUT}; however, in order to
evaluate the transformation from Brill's time coordinate $\tau$ to our
time coordinate $t$ one has to compute an integral which in general
cannot be represented in closed form. In any case, in order to keep
the expressions below as short as possible, the other three distinct
non-vanishing functions of the background
${\mathring{L}}\indices{_1_1}={\mathring{L}}\indices{_2_2}$,
$\mathring{E}^*_{11}$ and $\mathring{B}^*_{11}$ are not written out in
terms of $a_1(t)$, $a_3(t)$ and their derivatives in the following.

Note that for the full evolution equations with our choice of gauge
and frame, the $1$-$2$- and $2$-$3$-components of all covariant
$2$-tensor fields in the set of unknowns are identically zero for all
times at $\rho_1=0$. Also some of the components of the connection
coefficients vanish there. Hence, we do not need to consider their
evolution equations. Further we restrict our attention now to the
evolution equations of the electric and magnetic parts of the rescaled
Weyl tensor (the Bianchi system) since those describe the evolution of
curvature. Only where necessary, also the other equations are taken
into account. With $F(\chi):=-2\cot\chi$, the linearized evolution
equations take the following form 
\newcommand{\Termbrace}[2]{\underbrace{#1}_{\mathrm{Term\,#2}}\,\,}

\numparts
\begin{eqnarray}
  \label{eq:lin_evol_E22}
  &\eqalign{%
    \partial_t{{\hat E^*_{22}}}=&
    \Termbrace{- \frac{1}{2} {a_1}\left(2{\partial_\chi }
        +{F(\chi)}\right){\hat B}_{13}}{IV}
    \Termbrace{-\frac{{a_1^2} \hat B^*_{22}}{{a_3}}}{II}
    \Termbrace{+\left(\frac{{a_1^\prime}}{{a_1}}
        +\frac{2 {a_3^\prime}}{{a_3}}\right)\hat E^*_{22}}{I}\cr
    &\Termbrace{+\frac{3}{2}\mathring{B}^*_{11}\Big({\hat\Gamma}\indices{_1^2_3}
      +{\hat\Gamma}\indices{_2^1_3}\Big)
      +\frac{3}{2}\mathring{E}^*_{11} \Big(\hat\chi_{11}
      -\hat\chi_{22}\Big)}{V},}\\
  \label{eq:lin_evol_B22}
  &\eqalign{%
    \partial_t{{\hat B^*_{22}}}=&
    \Termbrace{\frac{1}{2}{a_1}\left(2{\partial_\chi }
        +{F(\chi)}\right) {\hat E}_{13}}{IV}
    \Termbrace{+\frac{{a_1^2} \hat E^*_{22}}{{a_3}}}{II}
    \Termbrace{+\left(\frac{{a_1^\prime}}{{a_1}}
        +\frac{2 {a_3^\prime}}{{a_3}}\right)\hat B^*_{22}}{I}\cr
    &\Termbrace{-\frac{3}{2}\mathring{E}^*_{11} \Big({\hat\Gamma}\indices{_1^2_3}
      +{\hat\Gamma}\indices{_2^1_3}\Big)
      +\frac{3}{2} \mathring{B}^*_{11} \Big(\hat\chi_{11}
      -\hat\chi_{22}\Big)}{V},}
\end{eqnarray}
\endnumparts
\numparts
\begin{eqnarray}
  \label{eq:lin_evol_E13}
  &\eqalign{%
    \partial_t{{{\hat E}_{13}}}=&
    \Termbrace{{3} {a_1} {\partial_\chi }\hat B^*_{11}}{III}
    \Termbrace{+ {a_1} \left({\partial_\chi}
        -{F(\chi)}\right)\hat B^*_{22}}{IV}\cr
    &\Termbrace{+\frac{2 {a_1^2} {\hat B}_{13}}{{a_3}}}{II}
    \Termbrace{+\Big(\frac{5 {a_1^\prime}}{2 {a_1}}
      +\frac{{a_3^\prime}}{2{a_3}}\Big) {\hat E}_{13}}{I}
    \Termbrace{-\frac{3}{2} \mathring{B}^*_{11} {\hat\Gamma}\indices{_3^2_3}
      -\frac{3}{2} \mathring{E}^*_{11} \hat\chi_{13}}{V},}\\
  \label{eq:lin_evol_B13}
  &\eqalign{%
    \partial_t{{{\hat B}_{13}}}=& 
    \Termbrace{-{3} {a_1} {\partial_\chi }{\hat E^*_{11}}}{III}
    \Termbrace{- {a_1}\left({\partial_\chi}
        -{F(\chi)}\right)\hat E^*_{22}}{IV}\cr
    &\Termbrace{-\frac{2 {a_1^2} {\hat E}_{13}}{{a_3}}}{II}
    \Termbrace{+\Big(\frac{5 {a_1^\prime}}{2 {a_1}}
      +\frac{{a_3^\prime}}{2 {a_3}}\Big){\hat B}_{13}}{I}
    \Termbrace{+\frac{3}{2} \mathring{E}^*_{11} {\hat\Gamma}\indices{_3^2_3}
      -\frac{3}{2}\mathring{B}^*_{11} \hat\chi_{13}}{V},}
\end{eqnarray}
\endnumparts
\numparts
\begin{eqnarray}
  \label{eq:lin_evol_E11}
  &\eqalign{%
    \partial_t{{{\hat E^*_{11}}}}=&
    \Termbrace{-\frac{1}{2}a_1\left(2{\partial_\chi }-
        {F(\chi)}\right){\hat B}_{13}}{III}
    \Termbrace{+\frac{3 {a_1^2} \hat B^*_{11}}{{a_3}}}{II}
    \Termbrace{+\frac{3 {a_1^\prime}{\hat E^*_{11}}}{{a_1}}}{I}\cr
    &\Termbrace{
      -\frac{3}{2}\mathring{B}^*_{11}\Big({\hat\Gamma}\indices{_1^2_3}
      -{\hat\Gamma}\indices{_2^1_3}\Big)
      -\frac{3}{2}\mathring{E}^*_{11} \Big(\hat\chi_{11}
      +\hat\chi_{22}\Big)}{V},}\\
  \label{eq:lin_evol_B11}
  &\eqalign{%
    \partial_t{{\hat B^*_{11}}} =& 
    \Termbrace{\frac{1}{2}a_1\left(2{\partial_\chi }
        - {F(\chi)}\right){\hat E}_{13}}{III}
    \Termbrace{-\frac{3 {a_1^2}{\hat E^*_{11}}}{{a_3}}}{II}
    \Termbrace{+\frac{3 {a_1^\prime} \hat B^*_{11}}{{a_1}}}{I}\cr
    &\Termbrace{
      +\frac{3}{2} \mathring{E}^*_{11} \Big({\hat\Gamma}\indices{_1^2_3}
      -{\hat\Gamma}\indices{_2^1_3}\Big)
      -\frac{3}{2}\mathring{B}^*_{11} \Big(\hat\chi_{11}
      +\hat\chi_{22}\Big)}{V}.}
\end{eqnarray}
\endnumparts

 
In order to avoid confusions we point out that studying solutions of
these equations, which originate in a linearization around a \LTN
spacetime, is fundamentally different from considering the
spin-$2$ system, i.e.\ the Bianchi system, on the same \LTN
background. Both lead to a similar, but distinct set of equations. In
the latter case, which is not of interest here, certain consistency
conditions \cite{penroserindler} must be satisfied, otherwise no
solutions need to exist at all.

\subsubsection{Dynamical effects at early times}
\begin{figure}[t]
  \begin{minipage}{0.49\linewidth} 
    \centering
    \includegraphics[width=\textwidth]{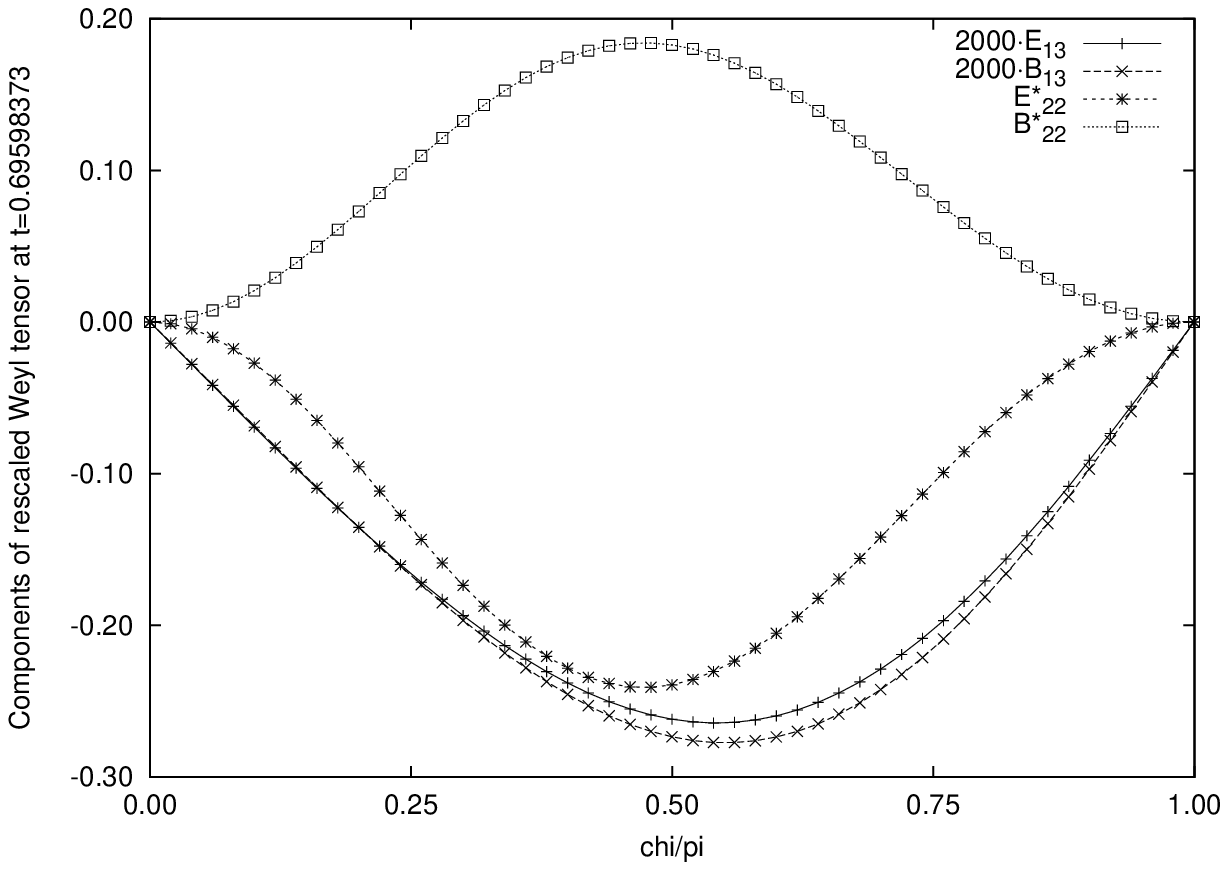}
    \caption{Spatial dependence at $t=0.69598373$ for $\eps=10^{-4}$}
    \label{fig:extremamove}
  \end{minipage}
  \begin{minipage}{0.49\linewidth} 
    \centering
    \includegraphics[width=\textwidth]{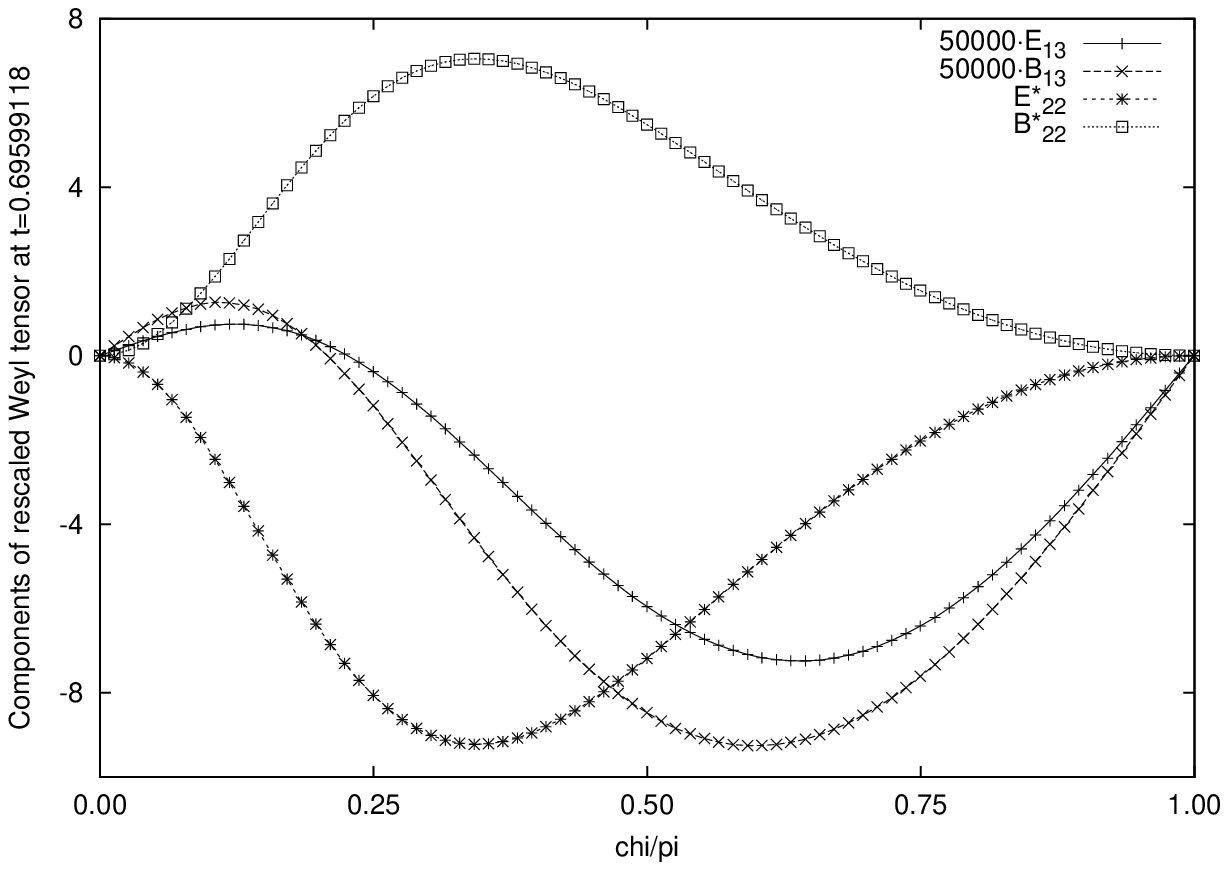}
    \caption{Spatial dependence at $t=0.69599118$ for $\eps=10^{-4}$}
    \label{fig:ausbeulen}
  \end{minipage}
\end{figure}

\paragraph{Oscillatory phase}
In this earliest phase in the past of \scrip, the ``evolution makes a
decision on the signs of the quantities'' by performing a couple of
oscillations before going over to a more or less monotonic
behavior. Consider the linearized evolution equations
\Eqref{eq:lin_evol_E22} and \Eqref{eq:lin_evol_B22} first. We know
from our choice of initial data that both $\hat E^*_{22}$ and $\hat
B^*_{22}$ are zero at $t=0$. If this is true at some other time, the
equations imply that also their time derivatives are zero, if
additionally $\hat \chi_{11}=\hat \chi_{22}$,
${\hat\Gamma}\indices{_1^2_3}=-{\hat\Gamma}\indices{_2^1_3}$ (term V)
and $\hat B_{13}$ and $\hat E_{13}$ are proportional to $\sin\chi$
(term IV); note here that $(2\partial_\chi+F(\chi))\sin\chi=0$. This
is the case initially. Checking the linearized evolution equations for
$\hat\chi_{11}-\hat \chi_{22}$ and
${\hat\Gamma}\indices{_1^2_3}+{\hat\Gamma}\indices{_2^1_3}$, which are
not printed here, one finds that also their time derivatives are zero
in that case. Further, due to the linearity of the equations and due
to the required behavior at the boundaries, $\hat E_{13}$ and $\hat
B_{13}$ must be proportional to $\sin\chi$ for all times. In summary,
the linearized evolution equations imply that for these initial data,
$\hat E^*_{22}$ and $\hat B^*_{22}$ are zero identically for all
times. Nevertheless, we find numerically that $\hat E^*_{22}$ and
$\hat B^*_{22}$ are non-zero for $t>0$. In fact, they grow so strongly
that they become dominant after a short time. Hence a discussion which
is exclusively based on the linearized equations with these initial
data would miss this important character of the solution. This fact is
related to the results in \cite{Arm79} which however restricts to
$\lambda=0$.  A simple nonlinear mechanism driving $\hat E^*_{22}$
and $\hat B^*_{22}$ away from zero is the following. If due to
nonlinear dynamics there is a small higher order component, say,
\begin{equation}
  \label{eq:nonlinear13}
  \hat B_{13}(t,\chi)=\hat B^{(0)}_{13}(t)\sin\chi
  +\hat B^{(1)}_{13}(t)\sin 2\chi
\end{equation}
with $\hat B^{(1)}_{13}\ll \hat B^{(0)}_{13}$, then term IV in
\Eqref{eq:lin_evol_E22} yields
\[(2\partial_\chi+F(\chi))\hat B_{13}=-4 B^{(1)}_{13}\sin^2\chi.\]
Indeed we find numerically that the early dynamics can be described
very well as
\begin{equation}
  \label{eq:22*_earlyprofile}
  \hat E^*_{22}(t,\chi)=\hat E^{*(0)}_{22}(t)\sin^2\chi;
\end{equation}
the same holds for the magnetic component. Another conceivable lowest
order contribution to $\hat E_{13}$ and $\hat B_{13}$ is one which is
proportional to $\cos\chi$, but then according to the boundary
conditions \Eqref{eq:BCs}, there must also be a term proportional to
$\cos 3\chi$ which is of higher order and can be neglected so far.
Another possibility consistent with the boundary conditions is of the
form $1-\cos 2\chi$. However then it turns out that term IV in
\Eqref{eq:lin_evol_E22} and \Eqref{eq:lin_evol_B22} violates
\Eqref{eq:BCs2}, i.e.\ the boundary conditions implied by the Bianchi
constraints. Hence \Eqref{eq:nonlinear13} is the lowest order
explanation for \Eqref{eq:22*_earlyprofile}. In any case, as soon as
$\hat E^*_{22}$ and $\hat B^*_{22}$ have been driven away from zero,
the linearized equations govern their evolution, in particular terms I
and II, see below. term IV does not become significant before $\hat
B^{(1)}_{13}$ in \Eqref{eq:nonlinear13} has become large enough in
comparison to $\hat B^{(0)}_{13}$; the same for the electric
component. At early times, we also find that terms V can be
neglected. One explanation is the following. When $t$ is small, the
only significant sources in the linearized evolution equations for the
quantities $\hat \chi_{11}-\hat \chi_{22}$ and
${\hat\Gamma}\indices{_1^2_3}+{\hat\Gamma}\indices{_2^1_3}$ are the
quantities $\hat E^*_{22}$ and $\hat B^*_{22}$ which are however very
small. Hence these quantities are even smaller. Now, since
$\mathring{E}^*_{11}$ and $\mathring{B}^*_{11}$ are not larger than
order unity for all times, this implies that term~V can be neglected
for small $t$.  What is the effect of terms~I~and~II? At early times
$a_1$ and $a_3$ differ only slightly, further $a_b^\prime/a_b\sim 1$
($b=1,3$) due to the choice of initial data, and thus term II can be
of the same order of magnitude as term I. Indeed, the sign structure
of these terms implies oscillations of the amplitudes $\hat
E^{*(0)}_{22}(t)$ and $\hat B^{*(0)}_{22}(t)$ in time. Now, both of
$a_1$ and $a_3$ are monotonically increasing, but $a_3$ eventually
stronger than $a_1$.  These oscillations stop when $a_3$ becomes so
big in comparison to $a_1$ that term II cannot have an significant
effect anymore; note that in particular $a_3^\prime/a_3$ is strongly
monotonically increasing as being proportional to
$\mathring{\chi}_{33}$. The oscillations die out relative early, and
in our notation above $\hat E^{*(0)}_{22}(t)<0$ and $\hat
B^{*(0)}_{22}(t)>0$ from then on. Then, term~I leads to strong growth
of the absolute values of the amplitudes. We find numerically, that
$|\hat E^{*(0)}_{22}(t)|>|\hat B^{*(0)}_{22}(t)|$ for all following
early times so that there is a net effect for the Kretschmann scalar
\Eqref{eq:physKretschConfQuant}.  Later in the evolution, this growth
becomes dominant and eventually seems to lead to blow up of the
Kretschmann scalar. In \Figref{fig:extremamove}, we see a plot of the
spatial distribution of these components at a time for which our
discussion so far is still valid. However, first signs of further
nonlinear effects become significant which we will discuss later.

Next consider the linearized evolution equations
\Eqref{eq:lin_evol_E13} and \Eqref{eq:lin_evol_B13} of $\hat E_{13}$
and $\hat B_{13}$. The early evolution takes place in a similar way as
before. The main difference is now however that the initial data of
$\hat E_{13}$ is non-vanishing and hence the linear effects described
by these equations dominate the nonlinear ones at the beginning.  In
the same way as above, terms I and II are dominant with a correction
in the oscillatory structure by the additional term III. Again,
oscillations stop when term~II becomes insignificant in comparison to
term~I. The influence of term V is different than before. Namely, by
checking the signs and magnitudes of ${\hat\Gamma}\indices{_3^2_3}$
and $\hat\chi_{13}$ numerically, term V can be seen to become
significant a bit later in the evolution.  Then it decelerates the
growth of $\hat E_{13}$ and $\hat B_{13}$, and so they do not blow up
as quickly as $\hat E^{*}_{22}$ and $\hat B^{*}_{22}$ above. The
significance of term IV becomes obvious later. For the whole early
time, $\hat E_{13}$ and $\hat B_{13}$ can be described very well as
proportional to $\sin\chi$, and we find numerically that their
amplitudes are negative from the time on at which oscillations stop;
consider again \Figref{fig:extremamove}. In particular, the plot shows
nicely the different rates of growth for the components discussed so
far; compare also to \Figref{fig:ausbeulen}.

The discussion of $\hat E^{*}_{11}$ and $\hat B^{*}_{11}$ and their
linearized evolution equations \Eqref{eq:lin_evol_E11} and
\Eqref{eq:lin_evol_B11} is very similar. Again term I and II are
dominant at early times and lead to oscillations in time with
corrections by term III. Eventually, after these have stopped, term V
decelerates the growth similarly as in the $1$-$3$-case so that they
grow much less than in particular $\hat E^{*}_{22}$ and $\hat
B^{*}_{22}$. Both $\hat E^{*}_{11}$ and $\hat B^{*}_{11}$ can be
described well as proportional to $\cos\chi$ with eventually positive
amplitudes.

For understanding the limit $\eps\rightarrow 0$, one should be worried
whether the relevant terms in the equations are of different relative
magnitude for different values of \eps. If yes, there would be a
dependence of the character of the early oscillations on \eps, and
this could lead to a different sign structure of the solution
eventually. Most likely this would mean that the late time behavior,
i.e.\ close to the critical time, depends strongly on \eps. However,
the structure of the linearized evolution equation and the initial
data are such, that the amplitudes of the dominant terms at early times
have a consistent scaling with \eps. Hence the early oscillations are,
apart from this scaling, independent of \eps. In detail, the
dependence of the initial data on \eps suggests \numparts
\begin{equation}
  \label{eq:scaling}
  \hat E^*_{11}, \hat B^*_{11}, \hat E_{13}, \hat B_{13}\sim\eps,
\end{equation}
which is compatible with the equations. Further it is reasonable to expect
\begin{equation}
  \label{eq:scaling2}
  \hat E^*_{22}, \hat B^*_{22}\sim\eps^2.
\end{equation}
\endnumparts
Indeed, one can check numerically that this scaling holds true for
early times, but we will not show this here. As it can be expected,
the behavior at later times cannot be described in this simple way
anymore.

\paragraph{Intermediate nonlinear phase}
We proceed with the discussion of that phase of the evolution in which
nonlinear effects take over.
In that phase, the evolution forms the spatial shapes of the unknowns,
and leading order effects are the following. As we have said before,
we find \Eqref{eq:22*_earlyprofile} for early times induced by the
higher order component in \Eqref{eq:nonlinear13}; similarly for the
magnetic component. From the sign structure of $\hat E^{*}_{22}$ and
$\hat B^{*}_{22}$ we can read off that the sign of this higher order
component must be such that the spatial extrema of $\hat E_{13}$ and
$\hat B_{13}$ slowly move to the south pole. Further we find that term
IV in \Eqref{eq:lin_evol_E13} and \Eqref{eq:lin_evol_B13} is then also
proportional to $\sin 2\chi$ with the ``right'' signs, hence this
effect is amplified.  In fact this is what we observe numerically, see
\Figref{fig:extremamove}.  Certainly, such nonlinear interactions are
also able to produce a component proportional to $\sin 3\chi$ or
$(\cos\chi-\cos 3\chi)$, recall the boundary conditions
\Eqref{eq:BCs}, for $\hat E_{13}$ and $\hat B_{13}$. For the latter
contribution, term IV in \Eqref{eq:lin_evol_E22} and
\Eqref{eq:lin_evol_B22} yields a violation of \Eqref{eq:BCs2}, hence
must not be present. Consider $\sin 3\chi$. With this, term IV in
\Eqref{eq:lin_evol_E22} and \Eqref{eq:lin_evol_B22} obtains a
component proportional to $\cos\chi\sin^2\chi$. The effect is that
also the extrema of $\hat E^*_{22}$ and $\hat B^*_{22}$ start to move
slowly, however, with this simple description we are not able to
deduce the signs of that term and hence the direction of the
move. Numerically we see that these extrema move toward the north
pole in \Figref{fig:extremamove} and \ref{fig:ausbeulen}.  Now,
considering again term IV in \Eqref{eq:lin_evol_E13} and
\Eqref{eq:lin_evol_B13} with this new component proportional to
$\cos\chi\sin^2\chi$, the quantities $\hat E_{13}$ and $\hat B_{13}$
obtain an additional contribution of the form $\sin 3\chi$, however,
with an amplitude opposite to the previous one; thus the nonlinear
interactions must dominate the linear ones here to keep this process
alive. We find numerically that the signs eventually become such that
the quantities $\hat E_{13}$ and $\hat B_{13}$ grow positively close
to the north pole, while the move of the extrema to the south pole is
decelerated. All this is visible in \Figref{fig:ausbeulen}. Indeed we
find eventually that the positive growth of $\hat E_{13}$ and $\hat
B_{13}$ at the north pole becomes much stronger than the dynamics at
the south pole.

When comparing different values of \eps, one finds that the smaller
$\eps>0$, the later, but eventually the more rapid, all these
nonlinear effects take place.  One may certainly wonder what is
special about, say, the north pole in these spacetimes. This issue
becomes clear when one recalls that going from $\eps$ to $-\eps$ is
equivalent to the interchange of the north and south pole.

\subsubsection{Universality and the late time behavior}
\begin{figure}[t]
  \begin{minipage}{0.49\linewidth}      
    \includegraphics[width=\textwidth]{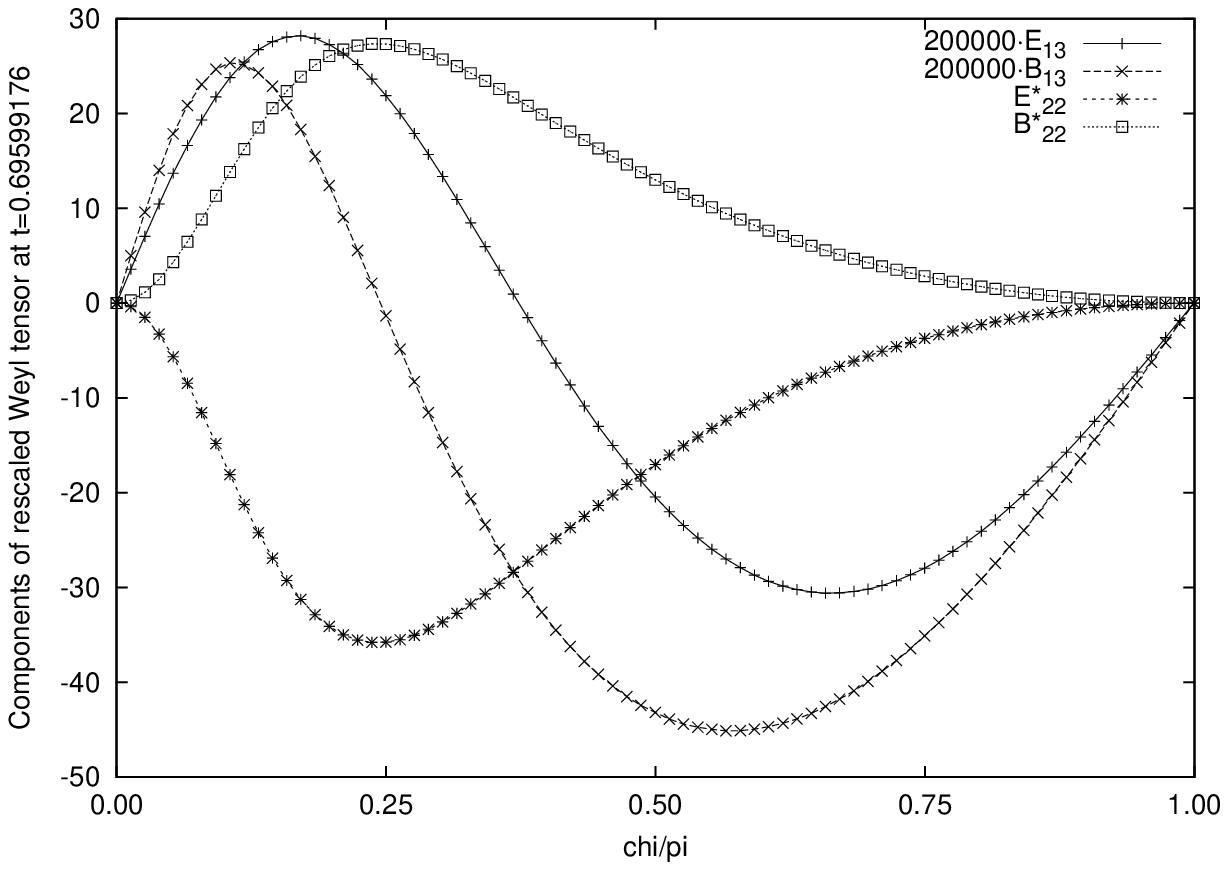}
    \caption{Spatial dependence at $t=0.69599176$ for $\eps=10^{-4}$}
    \label{eq:latetime1}
  \end{minipage}
  \begin{minipage}{0.49\linewidth}      
    \includegraphics[width=\textwidth]{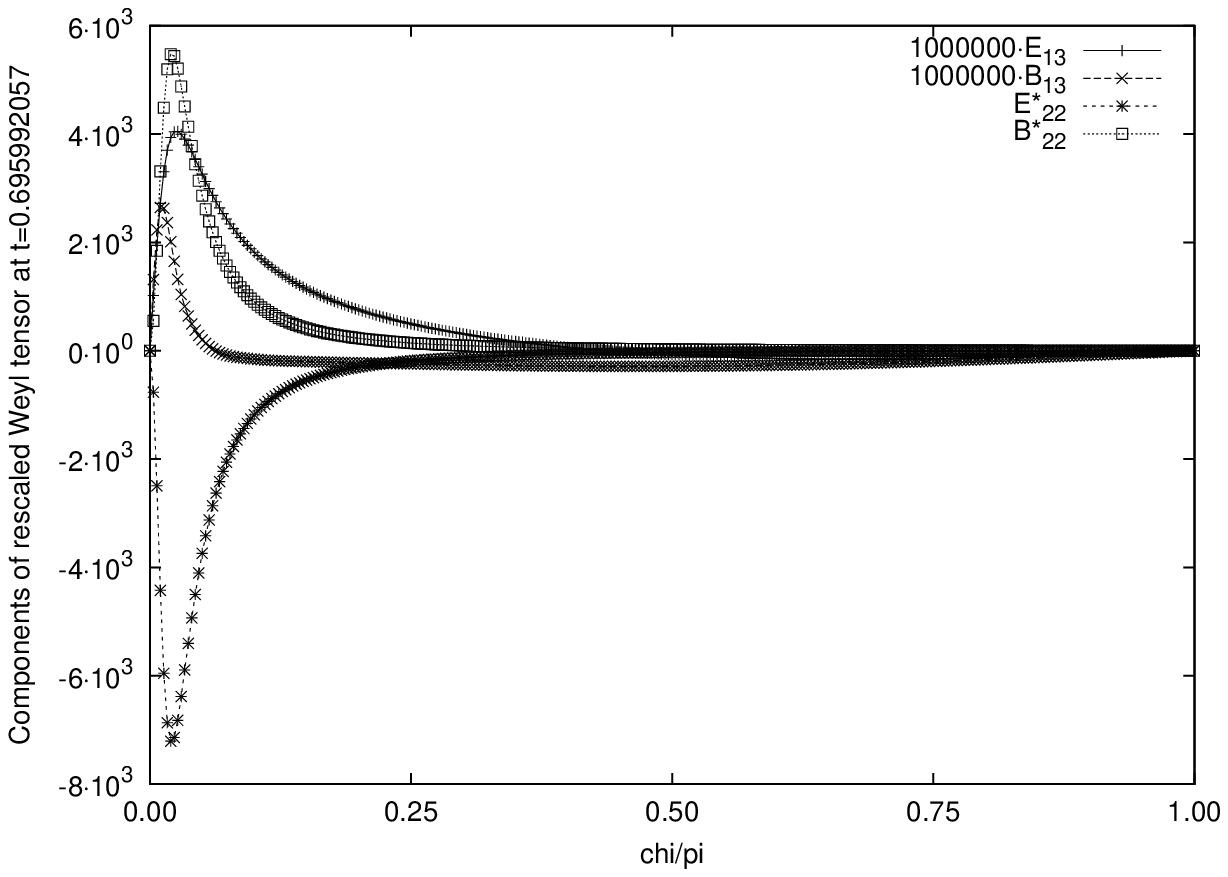}
    \caption{Spatial dependence at $t=0.69599206$ for $\eps=10^{-4}$}
  \end{minipage}\\
  \begin{minipage}{0.49\linewidth}      
    \includegraphics[width=\textwidth]{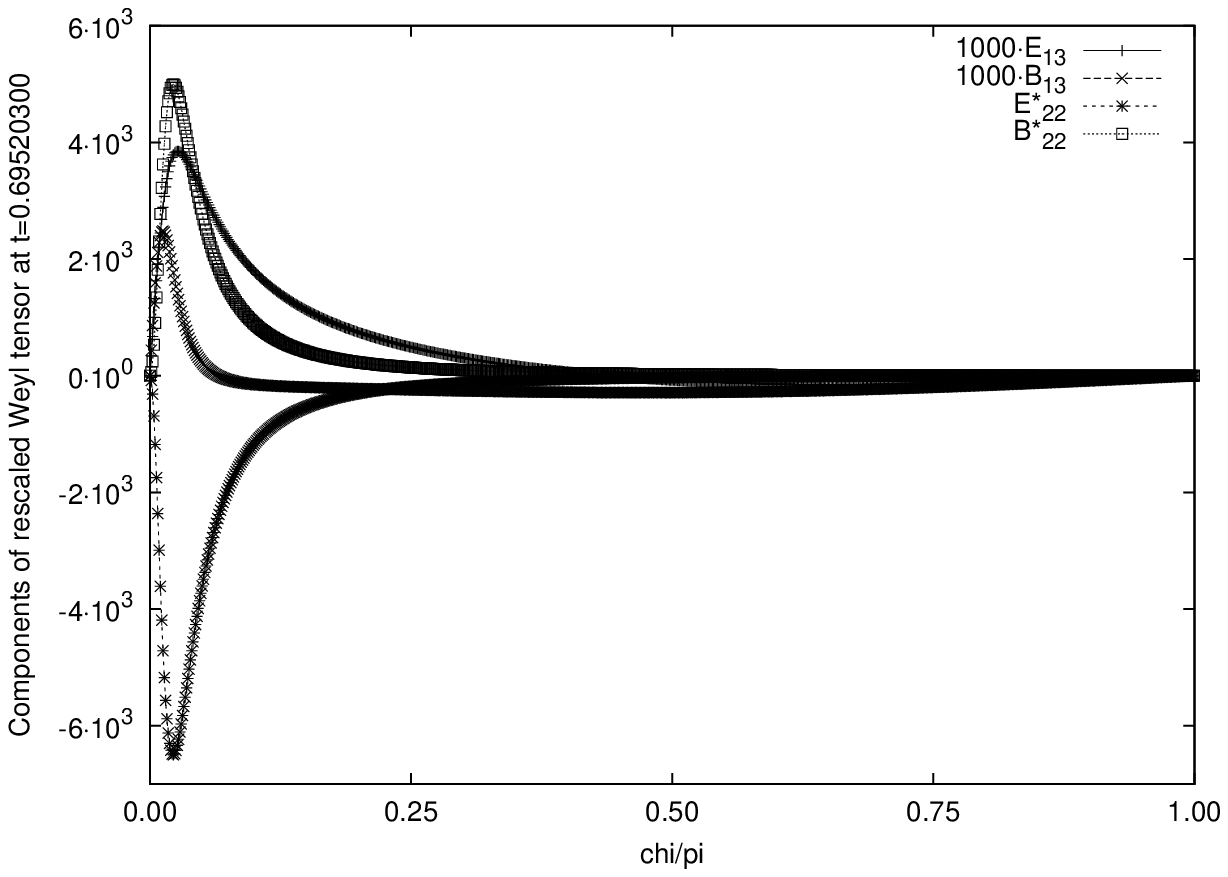}
    \caption{Spatial dependence at $t=0.69520300$ for $\eps=10^{-1}$}
  \end{minipage}
  \begin{minipage}{0.49\linewidth}      
    \includegraphics[width=\textwidth]{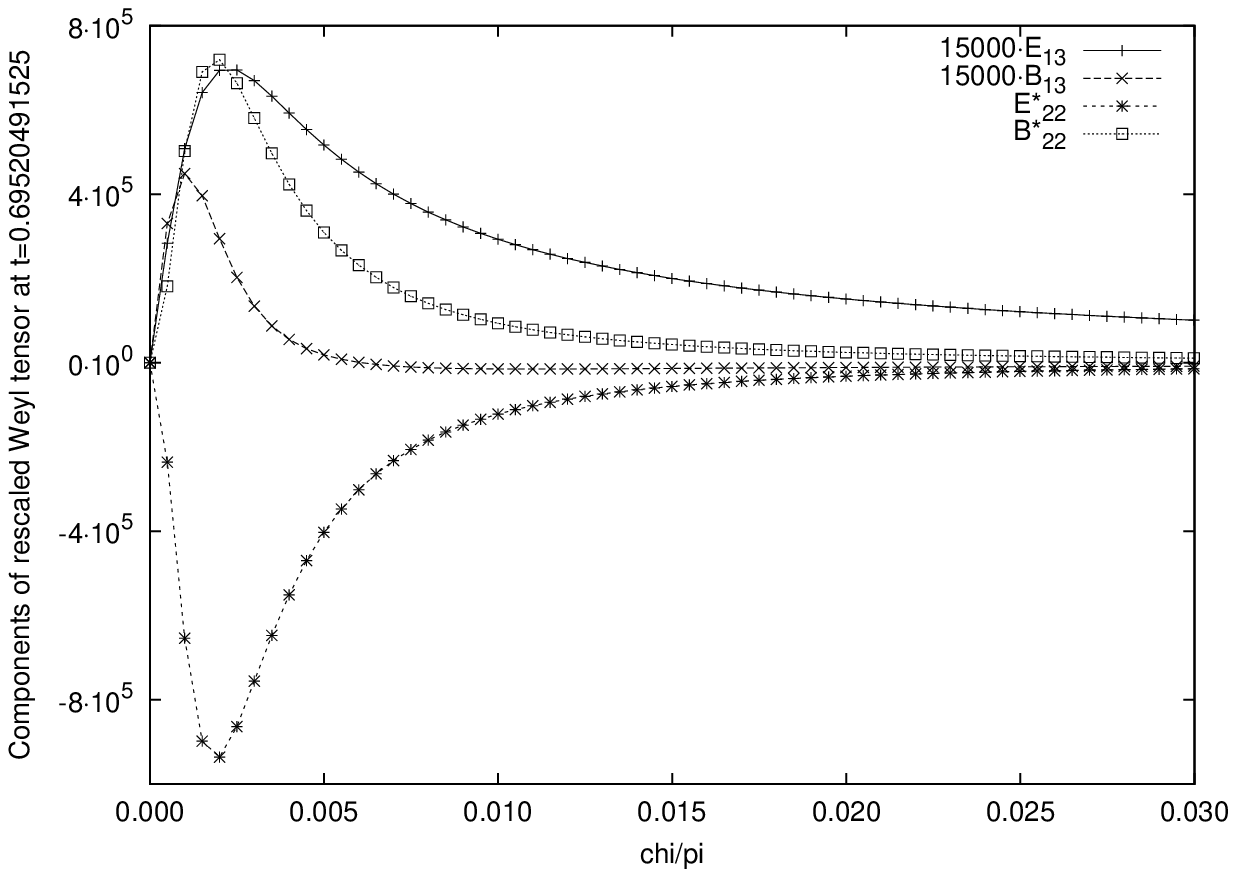}
    \caption{Spatial dependence at $t=0.69520492$ for $\eps=10^{-1}$}
    \label{eq:latetime4}
  \end{minipage}
\end{figure}
The longer the evolution proceeds, the higher is the order of the
significant nonlinear effects. We make no effort here to describe all
those. Nevertheless, we find numerically that the qualitative picture
described before does not change anymore for late times. Some of the
quantities tend to blow up and there is an even stronger concentration
of curvature close to $\chi=0$.  Further, the computations suggest
that qualitatively the same picture is true for all small values of
\eps. Certainly, nonlinear effects become significant at a later time
the smaller \eps is. In particular the magnitude of $E^{*}_{22}$ and
$B^{*}_{22}$ is smaller in relation to the other components at early
times according to the scaling given by \Eqref{eq:scaling} and
\Eqref{eq:scaling2}. But from some time on, and then on a shorter time
scale, these overtake and a ``universal'' qualitative picture is
obtained. All this is demonstrated in the numerical plots in
\Figref{eq:latetime1} to \ref{eq:latetime4}.  As already mentioned, we
see particularly strong blow up of $E^{*}_{22}$ and
$B^{*}_{22}$. Indeed, this seems to yields the main contribution to
the Kretschmann scalar at late times. According to the plots above,
$E_{13}$ and $B_{13}$ also grow at the north pole and maybe even blow
up eventually. However, this growth seems to be slower. Why does the
blow up of curvature seem to be confined to a small region around the
north pole? We can give an explanation for this below, and this is
most likely caused by our choice of foliation.

\begin{figure}[t]
  \begin{minipage}{0.49\linewidth} 
    \centering
    \includegraphics[width=\textwidth]{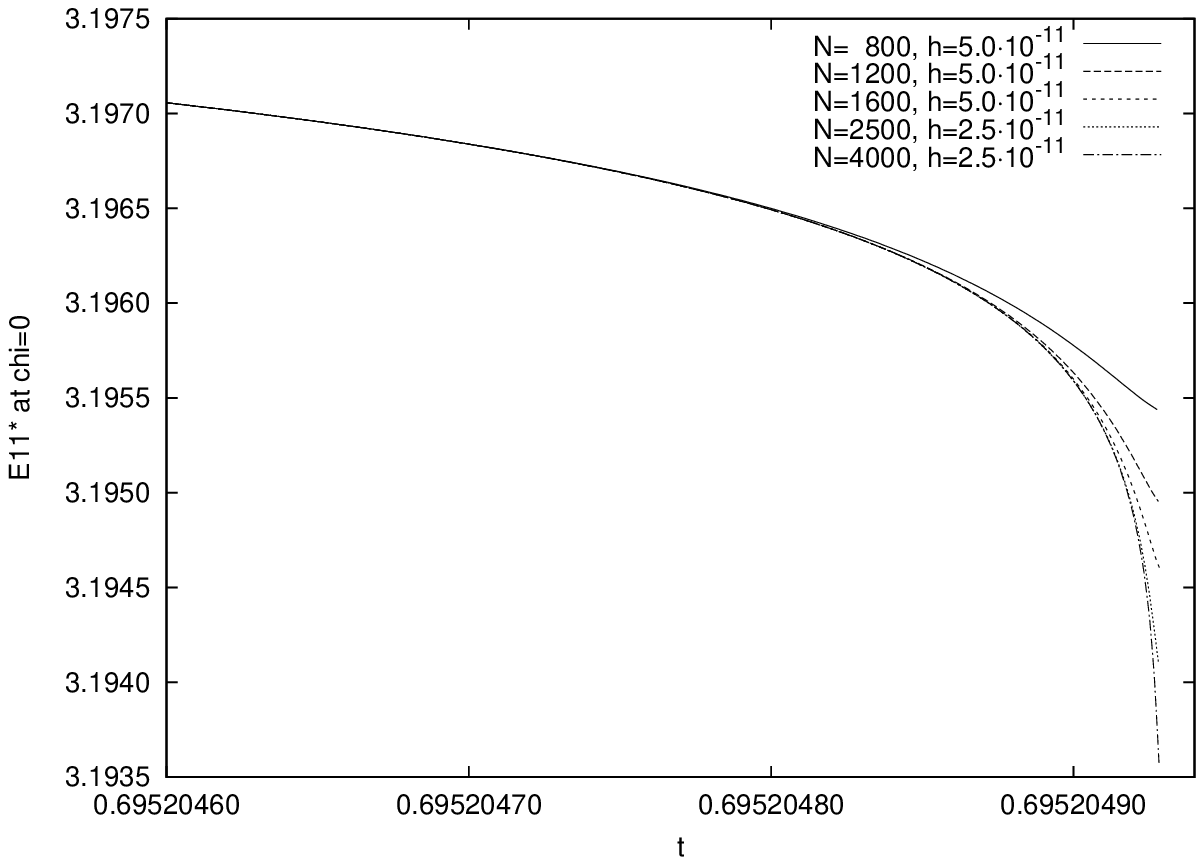}
    \caption{$E_{11}^*$ at $\chi=0$ for late times}
    \label{fig:behavior_pole_latest_times}
  \end{minipage}
  \begin{minipage}{0.49\linewidth} 
    \centering
    \includegraphics[width=\textwidth]{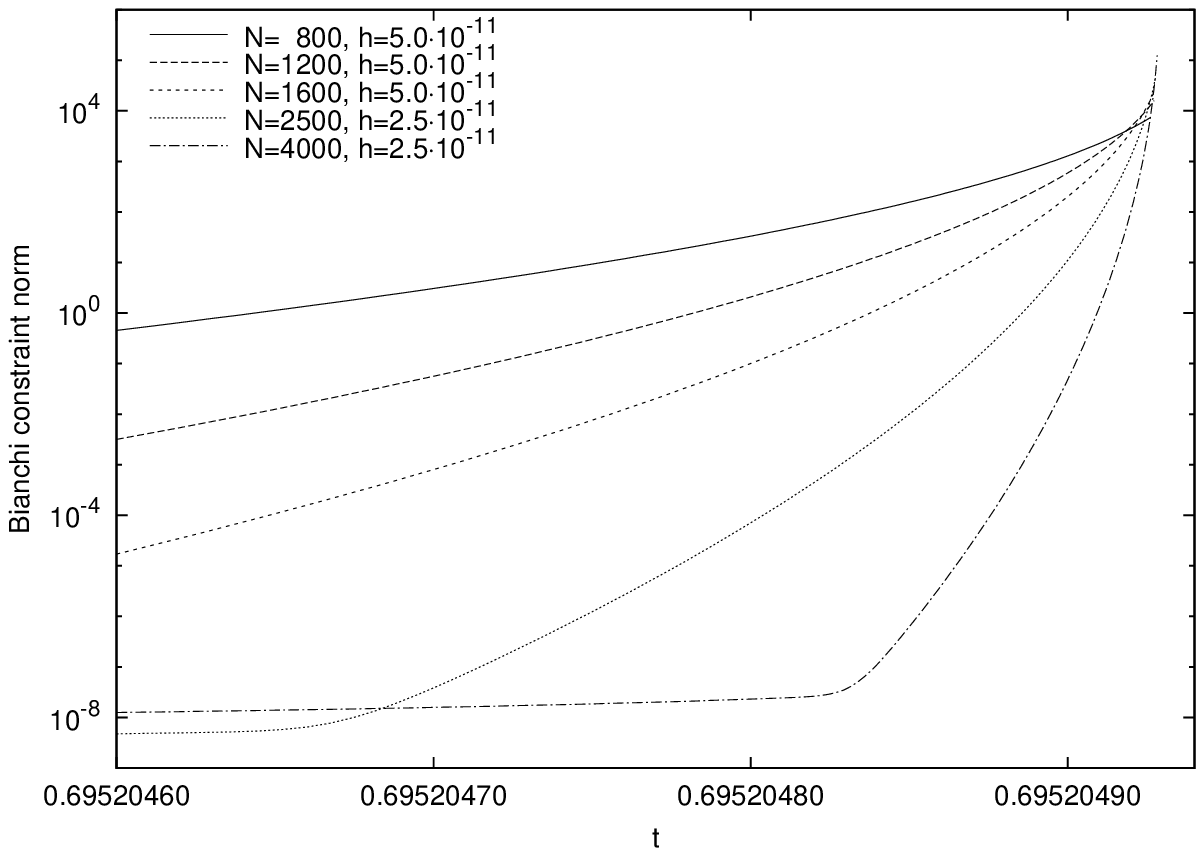}
    \caption{$\normconstr$ for late times}
    \label{fig:behavior_constraints_latest_times}
  \end{minipage}
\end{figure}
However, what happens exactly on the north pole? The only curvature
components which do not vanish there according to \Eqref{eq:BCs} are
$E^{*}_{11}$ and $B^{*}_{11}$ which we have said only little about so
far. It seems that, as long as we have evolved numerically so far,
those stay bounded and the dynamics is very slow. According to their
evolution equations, their behavior at $\chi=0$ depends strongly on
the behavior of the $\chi$-derivatives of $E_{13}$ and $B_{13}$
there. The numerical results suggest that those are positive and grow,
but relatively slowly. In fact, it seems that their growth is even
slower in comparison to the growth of $E^{*}_{22}$ and $B^{*}_{22}$
the smaller \eps is, see \Figref{eq:latetime1} to
\ref{eq:latetime4}. In total, it must be said that it is so far an
outstanding question if the curvature at $\chi=0$ stays bounded, in
particular for small values of \eps.  In order to demonstrate how
difficult these investigations are with the current numerical setup,
\Figref{fig:behavior_pole_latest_times} shows the behavior of
$E_{11}^*$ at $\chi=0$ for times shortly before the numerical
solutions blow up for various final numerical resolutions. Here $N$ is
the final number of spatial grid points in the $\chi$-direction and
$h$ is the final value of the size of the time step, i.e.\ equals
$h_{min}$ as given in \cite{beyer08:code}. In addition,
\Figref{fig:behavior_constraints_latest_times} shows the behavior of
the constraint violations $\normconstr$ defined in
\cite{beyer08:code}. This plot suggests that the numerical solution
converges and that the constraint error can be kept under control for
sufficiently high, in particular spatial resolutions at least
temporarily. Nevertheless, since on the one hand the demand for
resolution at those late times becomes so high, but on the other hand
the dynamics of $E_{11}^*$ (also $B_{11}^*$) at $\chi=0$ seems to be
so slow, we cannot answer the question about the behavior of curvature
on the north pole yet. This means that we are also not able yet to
exclude the possibility that a strong blow up of $E^{*}_{11}$ and
$B^{*}_{11}$ at very late times induces a drastic change of the whole
asymptotic dynamics.  
In any case, ideas how to optimize the numerical late time behavior
and the setup of our approach are discussed below.

\subsection{Further comments on the solutions}
\subsubsection{Area of the group orbits}
\begin{figure}[t]
  \begin{minipage}{0.49\linewidth} 
    \centering
    \includegraphics[width=\textwidth]{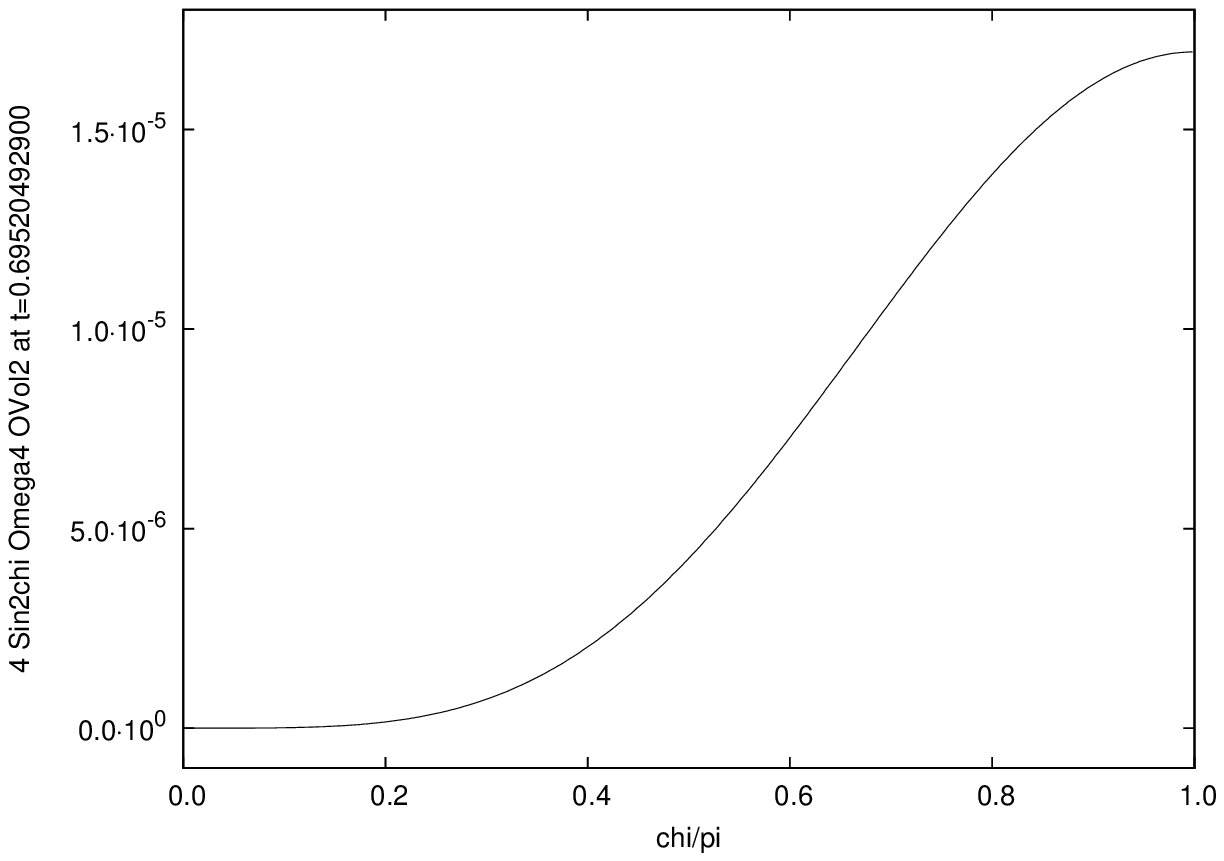}
    \caption{Rescaled orbit area at $t=0.69520493$ for $\eps=10^{-1}$}
    \label{fig:orbitarea1D}
  \end{minipage}
  \begin{minipage}{0.49\linewidth} 
    \centering
    \includegraphics[width=\textwidth]{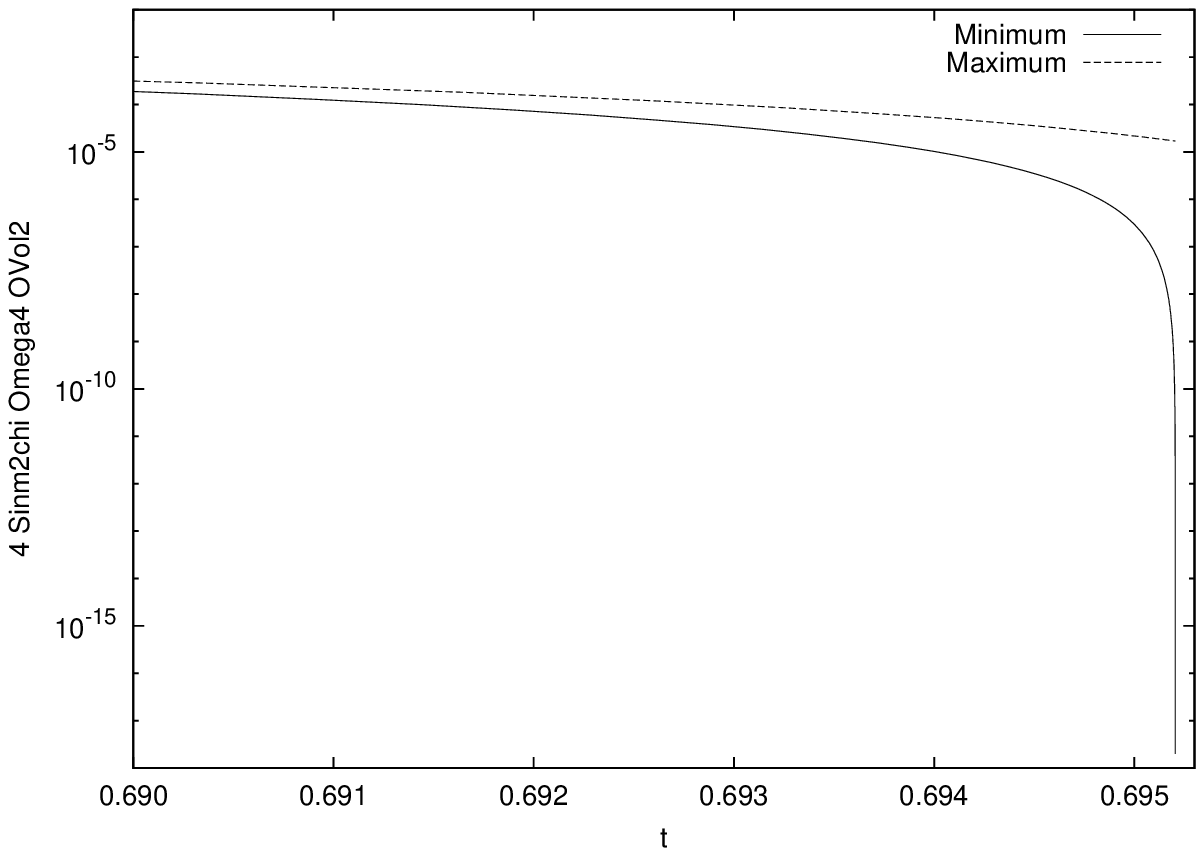}
    \caption{Extremal values of rescaled orbit area for $\eps=10^{-1}$
      on $t=const$-slices}
    \label{fig:orbitareaminmax}
  \end{minipage}
\end{figure}

In \Sectionref{sec:gowdy} we have introduced the notion of the
rescaled orbit area for the \S-Gowdy class and have stated that one
can believe that a relevant subset of the MGHDs can be covered by
areal foliations. In the \T-Gowdy case without cosmological constant
the singularity is attained generically exactly where the group orbit
area density vanishes. If we are willing to assume that this is true
also for our \S-Gowdy solutions with $\lambda>0$, then we can get a
feeling on how ``close'' the numerical evolutions have eventually
approached the ``singularity'' at the end of the numerical runs by
looking at the rescaled orbit area density. Certainly, our foliation
is not of areal type, hence this quantity is not constant on our
$t=\mathrm{const}$ slices, see \Figref{fig:orbitarea1D} for a typical
late time profile.
According to this interpretation, we see that our foliation has
approached the singularity further at the north pole than at the south
pole which explains why curvature blows up much more strongly
there. In fact, \Figref{fig:orbitareaminmax} implies that this
discrepancy between the poles, where the maximum and minimum
respectively of this rescaled quantity are always attained in our
case, becomes larger the further the evolution proceeds. Thus the
singularity is approached ``inhomogeneously'' in this sense in our
gauge. We just note that this is not surprising since many gauges
which have been used in the literature so far have such a
property. Hence one can expect that only special gauges are suitable
for our kinds of investigations, and it will be a matter of future
research to experiment with such other foliations, see
below. Concerning the question of SCC our gauge is thus not
appropriate enough. For other open questions, however, as that about
the appearance of Gowdy spikes, our computations have the potential to
yield new insights.

\subsubsection{Gowdy spikes}
From the results obtained in the \T-Gowdy case, one can expect spikes
in the sense of \Sectionref{sec:gowdy} to form under certain
conditions at isolated points away from the axes. The previous
discussion about the gauge however suggests that we cannot conclude
that the local curvature blow up close to the north pole is such a
spike.  Rather, it can be caused by the ``inhomogeneous'' approach to
the singularity. It is nevertheless possible that spikes show up later
in the evolution even in our gauge, but most likely they would be
difficult to distinguish from our gauge effects. However, from our
numerical data obtained in Levi-Civita conformal Gauß gauge we can --
in principle -- compute the geometrical quantities with respect to an
areal foliation until very ``close to the singularity'' in a small
spatial neighborhood of the north pole. With this it would be feasible
to study the possible appearance of spikes there. However, this has
not been done yet. In areal gauge, one often introduces a time
coordinate $\tau$ such that the rescaled orbit area is proportional to
$\e^{-\tau}$, at least for late times in the \S-case. Hence, a local
reconstruction of the areal foliation from the current numerical data
would reach Gowdy times of the order $\tau\approx 20$, which is in
fact quite impressive. Certainly, it is possible that our family of
solutions given by low \eps corresponds to ``low-velocity''
solutions. The concept of the asymptotic velocity is not defined here,
see \cite{Isenberg89,Stahl02} for the \S-case. In fact the relation of
this asymptotic quantity at the singularity and our definition of \eps
as a data parameter on \scrip is not easy to see. It sounds reasonable
that the larger \eps is, the higher the asymptotic value of the
velocity becomes. Now, in the \T-Gowdy case in vacuum with
$\lambda=0$, one finds that spikes occur only if the asymptotic
velocity exceeds a certain value. So we would expect spikes to form in
our class of spacetimes away from the axes only for large \eps. It is
an unexpected outcome of our numerical experiments that it seems
easier to compute solutions for larger \eps than for small values, see
below. Hence there is hope that spikes can be studied with the current
numerical setup. This is planned as a future research topic, but in
this paper we are rather concerned with the limit $\eps\rightarrow
0$. Nevertheless, we would like to note that our numerical method can
be expected to be able to resolve Gowdy spikes in principle because of
our experience with numerical investigation of \T-Gowdy solutions in
\cite{beyer:PhD}. The question of the behavior of the solutions on the
axes is particularly interesting since, as discussed in
\cite{Stahl02}, although one can expect some discontinuous behavior at
the axes in a particular family of parametrizations of the metric, it
is not clear whether generic \S-Gowdy solutions have a true spike
there.  Thus it is remarkable that our numerical results suggest that
even at those relatively late Gowdy times, the curvature has not blown
up on the axes yet. However, this might of course also be a
consequence of the special choice of initial data.

\subsection{Limitations and prospects for future work}
Our approach of studying the properties of this class of solutions has
several limitations so far, some of which were already mentioned
above. Some of these limitations are caused by our non-rigorous
combination of numerical and heuristic analyzes, and are of principle
nature. Others, however, can be avoided more easily and are currently
work in progress. In fact we believe that the numerical method has not
been pushed to its limits yet.

In our investigations, we are not yet able to conclude finally on the
question of SCC in this class of spacetimes, although the numerical
results suggest so far consistently that SCC holds in the sense
of \Sectionref{sec:mainresults}. Nevertheless, we cannot yet deduce
the very late time behavior of curvature at the north pole, mainly,
but not exclusively, due to numerical problems. Also away from the
north pole, we are facing the problem that the foliation does not
proceed toward the ``singularity'' fast enough as was discussed
above. Concerning this gauge issue, we would hence like to try various
other gauge conditions in future work, for instance areal gauge, which
might however be problematic for $\lambda>0$ in general, but possibly
not here. Within the class of general conformal Gauß gauges there is
also freedom left which we have not exploited yet. Nevertheless, it is
expected that the description of the transport of spatial symmetries
is not trivial in general in this gauge. We would like to note that
there is also gauge freedom left within the Levi-Civita conformal Gauß
gauge, namely to perform an arbitrary conformal transformation of the
initial 3-metric and correspondingly of the other data components. It
has not been investigated yet if it would be possible to optimize the
foliation in this simple way. Other promising gauges are those of
harmonic type and constant mean curvature foliations since there is
some experience with these in the numerical relativity
community. However, for their use with the conformal field equations,
it must be investigated first if and how those gauges can be put into
the required form.

Concerning the numerical method, there are currently a number of
obstacles for driving the numerical solutions ``further to the
singularity''. Some of those are straight forward to solve, others are
not. The runs presented here each took a few hours on single processor
machines. It is of course no principle problem to let the runs
continue for a few days etc. But at the moment, we have a problem with
disk space since our data output routines are not yet optimized;
clearly, this problem is straight forward to fix. What are the main
reasons that it is numerically demanding to continue the runs? At the
north pole, the solutions develop large gradients, hence much spatial
resolution is needed there. So far in our code, resolution cannot be
increased locally, and thus increasing the spatial resolution to cope
with the demands at the north pole always goes hand in hand with a
waste of grid points around the south pole where relatively little
happens. Up to now, we have investigated only little on spatial
coordinate transformations which would put more spatial points close
to the left axis and less to the right axis without a direct change of
the numerical infrastructure. In any case, this trick will be studied
further in the future. It should be noted that our code is not yet
optimized technically for high spatial resolutions in general; in
particular we still do not use the Fast Fourier Transform (FFT)
\cite{FFT} yet but only partial summation \cite{boyd}. Also it may be
true that there are more optimal time integrator schemes than the
adaptive Runge Kutta schemes of our choice, comments on this can be
found in \cite{boyd}. Further, it might make sense to think about
parallelization of the code; this should be straight forward with some
publically available FFT libraries as for instance \cite{FFTW}. It is
certainly a justified question if a pseudospectral method like ours is
suitable at all for this class of problems. Namely, although these
methods are highly accurate for lower resolutions they might be too
slow for high resolutions. Thus it makes sense to also investigate
into other methods, for instance finite differencing methods, maybe
even with multipatch or mesh refinement. For instance the similar
numerical investigations of Gowdy spacetimes with spatial
\SoXSt-topology in \cite{garfinkle1999} were done with finite
differencing. Another non-related numerical issue shows up for small
values of \eps at late times. Then, the numerical noise caused by
round-off errors can become significant. Indeed, this was the main
reason why we did not decrease \eps further than $10^{-5}$. Maybe, the
only solution to this problem is to switch to quad-precision which is
possible for Intel compilers and Intel processors, but then the code
runs much slower. Further comments on our numerical infrastructure can
be found in \cite{beyer08:code}.

A further aspect one should keep in mind is the question about the
optimal formulation of Einstein's field equations for our purposes. It
is justified to ask if it is necessary to use the conformal field
equations for our applications at all. There are two motivations why
we decided for these equations. First, prescribing data on $\scrip$ is
simpler than in the case of Cauchy surfaces of the physical
spacetime. Second, we wish to study FAdS spacetimes since those are
motivated by current cosmological observations and, which is maybe
more important here, the MGHD of any data on \scrip is geodesically
complete to the future by construction. So SCC is obeyed at least to
the future. In contrast, prescribing data on a Cauchy surface of the
physical spacetime yields no a priori knowledge in general whether the
corresponding solution is future geodesically complete. Of course, one
might want to restrict to the question of SCC in the past direction as
a first step. Then, the conformal field equations can be avoided. It
is possibly an advantage to rely on the experience of the numerical
relativity community which is available for certain formulations of
EFE but not so much for the conformal field equations.  For instance
there exist methods to damp constraint violations etc. Indeed, we have
performed successful numerical tests with a well-adapted system other
than the CFE in the \T-Gowdy case in \cite{beyer:PhD} and would like
to work out a similar system for the \S-case.


\section{Summary}
In this work, we have presented investigations of the evolution
properties of a special class of spacetimes given by certain
perturbations of the \LTN solutions. By means of a combination of
numerical and heuristic analyzes, we were able to draw a qualitative
picture of the late time behavior of the solutions. The results
suggest that SCC holds in this family and we are able to give more
details on the dynamical processes which seem to give rise to
SCC. However, we are not yet able to make final
conclusions. Nevertheless, we identify some of the problems and
limitations of our current approach and give prospects for future
work. It is surprising that the issue of strong cosmic censorship and
of the BKL-conjecture is still open even in this relatively simple
class of \S-Gowdy spacetimes. Certainly the family of spacetimes
considered here is very special, and it is unclear what information we
yield both about the general SCC conjecture and about SCC in the class
of general \S-Gowdy spacetimes. We believe however that the results
presented here are promising steps in order to derive a consistent
picture on these questions in the near future.


\ack
This work was supported in part by the Göran Gustaffson foundation. We
would like to thank in particular Helmut Friedrich, David Garfinkle,
Alan Rendall and Hans Ringström for useful discussions, advices and
comments.


\vspace{1cm}

\bibliography{bibliography}

\begin{thebibliography}{10}

\bibitem{Anderson04}
M.T. Anderson.
\newblock Existence and stability of even dimensional asymptotically de
  {S}itter spaces.
\newblock {\em Annales Henri Poincare}, 6:801--820, 2005,
  \urlalt{http://arxiv.org/abs/gr-qc/0408072}{gr-qc/0408072}.

\bibitem{andersson04a}
L.~Andersson.
\newblock The global existence problem in general relativity.
\newblock In P.T. Chru\'sciel and H.~Friedrich, editors, {\em The Einstein
  Equations and the Large Scale Behavior of Gravitational Fields: 50 Years of
  the Cauchy Problem in General Relativity}, pages 71--120. Birkh\"auser,
  Basel, Switzerland; Boston, U.S.A., 2004.

\bibitem{galloway2002}
L.~Andersson and G.J. Galloway.
\newblock d{S}/{CFT} and spacetime topology.
\newblock {\em Adv. Theor. Math. Phys.}, 6:307--327, 2002,
  \urlalt{http://arxiv.org/abs/hep-th/0202161}{hep-th/0202161}.

\bibitem{Arm79}
J.~Arms and J.~Marsden.
\newblock The absence of {K}illing fields is necessary for linearization
  stability of {E}instein's equations.
\newblock {\em Indiana Univ. Math. J.}, 28:119--125, 1979.

\bibitem{belinskii70}
V.A. Belinskii, I.M. Khalatnikov, and E.M. Lifshitz.
\newblock Oscillatory approach to a singular point in the relativistic
  cosmology.
\newblock {\em Adv. Phys.}, 19:525--573, 1970.

\bibitem{belinskii82}
V.A. Belinskii, I.M. Khalatnikov, and E.M. Lifshitz.
\newblock A general solution of the {E}instein equations with a time
  singularity.
\newblock {\em Adv. Phys.}, 31:639--667, 1982.

\bibitem{beyer:PhD}
F.~Beyer.
\newblock {\em Asymptotics and singularities in cosmological models with
  positive cosmological constant}.
\newblock PhD thesis, Max Planck Institute for Gravitational Physics, 2007,
  \urlalt{http://www.arxiv.org/abs/0710.4297}{gr-qc/0710.4297}.

\bibitem{Anderson08}
F.~Beyer, 2008.
\newblock Private communication with M.T.~Anderson.

\bibitem{beyer08:code}
F.~Beyer.
\newblock A numerical approach for hyperbolic problems with spatial
  {$\S$}-topology.
\newblock preprint, 2008,
  \urlalt{http://arxiv.org/abs/0804.4222}{arXiv:0804.4222 [gr-qc]}.

\bibitem{boyd}
J.P. Boyd.
\newblock {\em Chebyshev and Fourier Spectral Methods}.
\newblock Dover Publications, Inc., 2nd edition, 2001.

\bibitem{Brill1978}
D.~{Brill} and F.~{Flaherty}.
\newblock {Maximizing properties of extremal surfaces in general relativity}.
\newblock {\em Annales de L'Institut Henri Poincare Section Physique
  Theorique}, 28:335--347, May 1978.

\bibitem{Chrusciel90}
P.~Chru\'{s}ciel, J.~Isenberg, and V.~Moncrief.
\newblock Strong cosmic censorship in polarized {G}owdy space-times.
\newblock {\em Class. Quant. Grav.}, 7:1671--1680, 1990.

\bibitem{chrusciel1990}
P.T. Chru\'{s}ciel.
\newblock On space-times with {$U(1)\times U(1)$} symmetric compact {C}auchy
  surfaces.
\newblock {\em Annals Phys.}, 202:100--150, 1990.

\bibitem{chrusciel93}
P.T. Chru\'{s}ciel and J.~Isenberg.
\newblock Nonisometric vacuum extensions of vacuum maximal globally hyperbolic
  spacetimes.
\newblock {\em Phys. Rev. D}, 48(4):1616--1628, Aug 1993.

\bibitem{FFT}
J.W. Cooley and J.W. Tukey.
\newblock An algorithm for the machine calculation of complex {F}ourier series.
\newblock {\em Math. Comput.}, 19:297--301, 1965.

\bibitem{DeSitter}
H.~Friedrich.
\newblock Existence and structure of past asymptotically simple solutions of
  {E}instein's field equations with positive cosmological constant.
\newblock {\em J. Geom. Phys.}, 3(1):101--117, 1986.

\bibitem{friedrich86}
H.~Friedrich.
\newblock On the existence of $n$-geodesically complete or future complete
  solutions of {E}instein's field equations with smooth asymptotic structure.
\newblock {\em Commun. Math. Phys.}, 107:587--609, 1986.

\bibitem{AntiDeSitter}
H.~Friedrich.
\newblock Einstein equations and conformal structure: Existence of anti-{D}e
  {S}itter-type spacetimes.
\newblock {\em J. Geom. Phys.}, 17:125--184, 1995.

\bibitem{Friedrich2002}
H.~Friedrich.
\newblock {\em The Conformal Structure of Spacetime: Geometry, Analysis,
  Numerics}, chapter "Conformal Einstein Evolution".
\newblock Lecture Notes in Physics. Springer, 2002.

\bibitem{FriedrichNagy}
H.~Friedrich and G.~Nagy.
\newblock The initial boundary value problem for {E}instein's vacuum field
  equations.
\newblock {\em Commun. Math. Phys.}, 201:619--655, 1999.

\bibitem{Friedrich98}
H.~Friedrich, I.~Racz, and R.M. Wald.
\newblock On the rigidity theorem for spacetimes with a stationary event
  horizon or a compact {C}auchy horizon.
\newblock {\em Commun. Math. Phys.}, 204:691--707, 1999,
  \urlalt{http://arxiv.org/abs/gr-qc/9811021}{gr-qc/9811021}.

\bibitem{ConfGeodesics}
H.~Friedrich and B.~G. Schmidt.
\newblock Conformal geodesics in general relativity.
\newblock {\em Proc. Roy. Soc. Lond. A}, 414:171--195, 1987.

\bibitem{FFTW}
M.~Frigo and S.G. Johnson.
\newblock {FFTW} {L}ibrary, \url{http://www.fftw.org/}.

\bibitem{garfinkle1999}
D.~Garfinkle.
\newblock Numerical simulations of {G}owdy spacetimes on {$S^{2}\times
  S^{1}\times R$}.
\newblock {\em Phys. Rev. D}, 60(10):104010, Oct 1999,
  \urlalt{http://arxiv.org/abs/gr-qc/9906019}{gr-qc/9906019}.

\bibitem{Gowdy73}
R.H. Gowdy.
\newblock Vacuum space-times with two parameter spacelike isometry groups and
  compact invariant hypersurfaces: {T}opologies and boundary conditions.
\newblock {\em Ann. Phys.}, 83:203--241, 1974.

\bibitem{hawking}
S.W. Hawking and G.F.R. Ellis.
\newblock {\em The large scale structure of space-time}.
\newblock Cambridge University Press, 1973.

\bibitem{Hawking82}
S.W. Hawking and I.L. Moss.
\newblock Supercooled phase transitions in the very early universe.
\newblock {\em Phys. Lett. B}, 110(1):35--38, 1982.

\bibitem{Intel}
Intel.
\newblock Intel {F}ortran {C}ompiler,
  \urlalt{http://www.intel.com/support/performancetools/fortran/}{http://www.i%
ntel.com/support/performancetools/}.

\bibitem{Isenberg89}
J.~Isenberg and V.~Moncrief.
\newblock Asymptotic behavior of the gravitational field and the nature of
  singularities in {G}owdy space-times.
\newblock {\em Ann. Phys.}, 199:84--122, 1990.

\bibitem{lifshitz63}
E.M. Lifshitz and I.M. Khalatnikov.
\newblock Investigations in relativistic cosmology.
\newblock {\em Adv. Phys.}, 12:185--249, 1963.

\bibitem{moncrief84}
V.~Moncrief.
\newblock The space of (generalized) {Taub-NUT} spacetimes.
\newblock {\em J. Geom. Phys.}, 1(1):107--130, 1984.

\bibitem{moncrief83}
V.~Moncrief and J.~Isenberg.
\newblock Symmetries of cosmological {C}auchy horizons.
\newblock {\em Commun. Math. Phys.}, 89:387--413, 1983.

\bibitem{NUT63}
E.~Newman, L.~Tamburino, and T.~Unti.
\newblock Empty-space generalization of the {S}chwarz\-schild metric.
\newblock {\em J. Math. Phys.}, 4(7):915--923, 1963.

\bibitem{penrose1963}
R.~Penrose.
\newblock Asymptotic properties of fields and space-time.
\newblock {\em PRL}, 10:66--68, 1963.

\bibitem{Penrose69}
R.~Penrose.
\newblock Gravitational collapse: The role of general relativity.
\newblock {\em Riv. Nuovo Cim.}, 1:252--276, 1969.

\bibitem{penrose1979}
R.~Penrose.
\newblock Singularities and time-asymmetry.
\newblock In S.W. Hawking and W.~Israel, editors, {\em General Relativity -- An
  Einstein Centenary Survey}. Cambridge University Press, 1979.

\bibitem{penroserindler}
R.~Penrose and W.~Rindler.
\newblock {\em Spinors and Space-Time}, volume~1.
\newblock Cambridge University Press, 1986.

\bibitem{numericalrecipes}
W.H. Press, S.A. Teukolsky, W.T. Vetterlin, and B.P. Flannery.
\newblock {\em Numerical Recipes in C}.
\newblock Cambridge University Press, 2nd edition, 1999.

\bibitem{Rendall05}
A.D. Rendall.
\newblock Theorems on existence and global dynamics for the {E}instein
  equations.
\newblock {\em Living Reviews in Relativity}, 8(6), 2005,
  \urlalt{http://www.livingreviews.org/lrr-2005-6}{lrr-2005-6}.

\bibitem{Ringstrom99}
H.~Ringström.
\newblock Curvature blow up in {B}ianchi {VIII} and {IX} vacuum spacetimes.
\newblock {\em Class. Quant. Grav.}, 17:713--731, 2000,
  \urlalt{http://arxiv.org/abs/gr-qc/9911115}{gr-qc/9911115}.

\bibitem{Ringstrom06}
H.~Ringström.
\newblock Existence of an asymptotic velocity and implications for the
  asymptotic behavior in the direction of the singularity in {$T^3$}-{G}owdy.
\newblock {\em Comm. Pure Appl. Math.}, 59(7):977--1041, 2006.

\bibitem{Ringstrom06b}
H.~Ringström.
\newblock Strong cosmic censorship in {$T^3$}-{G}owdy spacetimes.
\newblock preprint, accepted in Ann. of Math., 2006,
  \url{http://www.math.kth.se/~hansr/}.

\bibitem{Sanchez06}
A.G. Sanchez and C.M. Baugh.
\newblock Cosmological parameters 2006.
\newblock preprint, 2006,
  \urlalt{http://arxiv.org/abs/astro-ph/0612743}{astro-ph/0612743}.

\bibitem{Spergel06}
D.~N. Spergel et~al.
\newblock Three-year {W}ilkinson {M}icrowave {A}nisotropy {P}robe ({WMAP})
  observations: {I}mplications for cosmology.
\newblock {\em Astroph.\ J.\ Suppl.\ S.}, 170(2):377--408, 2007,
  \urlalt{http://arxiv.org/abs/astro-ph/0603449}{astro-ph/0603449}.

\bibitem{Stahl02}
F.~St{\aa}hl.
\newblock Fuchsian analysis of {$S^2\times S^1$ and $S^3$ Gowdy} spacetimes.
\newblock {\em Class. Quant. Grav.}, 19:4483--4504, 2002,
  \urlalt{http://arxiv.org/abs/gr-qc/0109011}{gr-qc/0109011}.

\bibitem{Taub51}
A.~H. Taub.
\newblock Empty space-times admitting a three parameter group of motions.
\newblock {\em Annals of Mathematics}, 53(3):472--490, May 1951.

\end{thebibliography}

\end{document}